\documentclass[lettersize,journal]{IEEEtran}
\usepackage{amsmath,amsfonts}
\usepackage{algorithmic}
\usepackage{algorithm}
\usepackage{array}
\usepackage[caption=false,font=normalsize,labelfont=sf,textfont=sf]{subfig}
\usepackage{textcomp}
\usepackage{stfloats}
\usepackage{url}
\usepackage{verbatim}
\usepackage{graphicx}
\usepackage{enumitem}
\usepackage{cite}
\usepackage{amsthm}
\usepackage{tabularx}
\usepackage{booktabs}   % 提供 \toprule \midrule \bottomrule
\usepackage{pdflscape}  % 提供 landscape 环境（PDF 自动旋转）
\usepackage{placeins}

\theoremstyle{plain}
\newtheorem{theorem}{Theorem}[section]        % Theorem 1.1, 1.2, ...
\theoremstyle{definition}
\newtheorem{definition}[theorem]{Definition}

\theoremstyle{remark}

\begin{document}

\newcommand{\SYSNAME}{EBCC}

\title{\SYSNAME: Enclave-Backed Confidential Containers via OCI-Compatible Runtime Integration}

\author{Di~Lu,~\IEEEmembership{Member,~IEEE,}
Qingwen Zhang,
Yujia Liu,
Xuewen~Dong,~\IEEEmembership{Member,~IEEE,}
Yulong~Shen,~\IEEEmembership{Member,~IEEE,}
Zhiquan~Liu, ~\IEEEmembership{Senior Member,~IEEE,}
and~Jianfeng~Ma,~\IEEEmembership{Member,~IEEE}
% <-this % stops a space
\thanks{\textbullet\ Di Lu, Qingwen Zhang, Yujia Liu, Xuewen Dong, and Yulong Shen are with the School of Computer Science and Technology, Xidian University, Xi’an, Shaanxi 710071, China, and also with the Shaanxi Key Laboratory of Network and System Security, Xi'an, Shaanxi 710071, China. 
E-mail: \{dlu, xwdong\}@xidian.edu.cn; 15319018420@163.com; 19829259628@163.com; ylshen@mail.xidian.edu.cn.\\
\textbullet\ Zhiquan Liu is with the College of Cyber Security, Jinan University, Guangzhou 510632, China. (E-mail: zqliu@vip.qq.com).\\
\textbullet\ Jianfeng Ma is with the School of Cyber Engineering, Shaanxi Key Lab of Network and System Security, Xidian University, Xi’an, China. E-mail: jfma@mail.xidian.edu.cn.}%
\thanks{Manuscript received April 19, 2021; revised August 16, 2021.}}

\markboth{Journal of \LaTeX\ Class Files,~Vol.~14, No.~8, August~2021}%
{Shell \MakeLowercase{\textit{et al.}}: A Sample Article Using IEEEtran.cls for IEEE Journals}


\maketitle

\begin{abstract}
Container runtimes provide a stable operational interface for deploying, monitoring, and controlling modern workloads, while trusted execution environments (TEEs) provide hardware-enforced isolation for sensitive computation. Existing confidential-container systems often rely on VM-backed deployment stacks or TEE-specific execution substrates, which can separate confidential execution from the conventional OCI runtime lifecycle. This paper presents \SYSNAME\ (\emph{Enclave-Backed Confidential Containers}), an OCI-compatible runtime architecture for managing composite confidential-computing workloads. \SYSNAME\ treats the REE-side anchor and TEE-side confidential stages as a single containerized confidential-computing composite, while keeping TEE-specific execution behind a backend adapter. The runtime preserves standard OCI lifecycle operations, maintains persistent per-instance state across multi-call runtime invocations, and materializes per-stage request, response, metadata, log, and evidence artifacts.

We implement \SYSNAME\ on a Keystone backend and evaluate its lifecycle behavior, performance overhead, footprint, and concurrent execution behavior. The results show that EBCC's cold-start overhead is mainly concentrated in the \textit{create} phase, while the \textit{start} phase remains small. Warm execution reduces median bring-up latency from about 2.01~s to 1.48~s under the light workload and from about 2.03~s to 1.51~s under the heavy workload. End-to-end experiments show that EBCC introduces additional latency over native Keystone, but the overhead corresponds to lifecycle mediation, request validation, EID allocation, backend dispatch, and evidence persistence. Under concurrent execution, EBCC completes all tested rounds successfully and sustains stable effective stage throughput. Cross-TEE case studies on SGX, TDX, and OP-TEE further show that the same lifecycle and stage abstraction can be mapped to enclave-style, VM-style, and embedded-style TEEs. These results indicate that EBCC can make TEE-backed execution manageable through an OCI-style lifecycle while keeping the protected-side TCB narrow.
\end{abstract}


\section{Introduction}\label{sec:intro}

Container runtimes have become the default substrate for deploying modern services, not only in cloud datacenters but increasingly at the edge. Their appeal largely comes from operational simplicity: developers package an application once, and operators rely on a stable toolchain---Docker, containerd, and OCI runtimes---to manage execution, logging, and lifecycle at scale. Meanwhile, a growing class of workloads demands stronger isolation than conventional OS-level containers can provide. Sensitive data processing, proprietary algorithms, and policy-constrained computation are routinely executed on machines outside the application owner's administrative domain. In these settings, compromise of the host OS or privileged runtime components can directly lead to data leakage or workload tampering.

Trusted execution environments (TEEs) offer a principled way to protect sensitive computation by isolating code and data even in the presence of a compromised host. This has motivated TEE-backed protection for containerized workloads, with the goal of combining hardware-enforced isolation with the operational convenience of containers. However, integrating TEEs into container deployment pipelines remains nontrivial. Existing systems often follow one of two directions. VM-backed confidential-container stacks run protected workloads inside confidential VMs, providing strong isolation but introducing a VM-granularity execution model and additional attestation/control-plane complexity~\cite{coco_ccs24_shiftedmoats,segarra2024serverlesscoco}. Enclave-centric systems, such as SCONE, Occlum, and Graphene-SGX, provide library-OS or compatibility-layer support for running applications inside process-level enclaves~\cite{scone_osdi16,occlum_asplos20,graphene_sgx_atc17}. These approaches are valuable, but they often separate confidential execution from the conventional OCI runtime lifecycle, or require TEE-aware launchers, execution substrates, and management paths.

This separation creates a practical systems problem. A real confidential-computing workload is often not a single enclave process. Instead, it is a composite execution: a normal-world component handles business logic, external interaction, I/O, and orchestration-facing behavior, while one or more protected stages are invoked on demand to process sensitive data. Managing only the protected component is therefore semantically incomplete, while moving the entire container runtime or orchestration logic into the TEE would unnecessarily enlarge the trusted computing base (TCB). What is missing is a runtime structure that preserves the standard OCI lifecycle for the composite workload, while keeping TEE-specific execution confined to a backend stage interface.

This paper presents \SYSNAME\ (\emph{Enclave-Backed Confidential Containers}), an OCI-compatible runtime architecture for managing such composite confidential-computing workloads. The central abstraction of \SYSNAME\ is the \emph{Containerized Confidential-Computing Composite} (C4), which treats the REE-side anchor and TEE-side confidential stages as one OCI-managed execution instance. The container manager still interacts with the workload through standard OCI entry points, such as \texttt{create}, \texttt{start}, \texttt{state}, \texttt{wait}, \texttt{kill}, and \texttt{delete}. Internally, however, \SYSNAME\ separates the instance-level lifecycle, identified by a composite instance identifier (CID), from stage-level confidential execution, where each protected invocation is represented by an enclave/stage identifier (EID). This separation allows the runtime to preserve container-compatible lifecycle behavior while invoking TEE-backed stages only when required by the anchor workflow.

To make this design operational, \SYSNAME\ maintains persistent host-side coordination state and materializes durable per-stage artifacts. The runtime records CID-level lifecycle state in files such as \texttt{state.json} and \texttt{session.json}; it also maintains request and response directories, EID allocation state, per-stage metadata, execution logs, and backend-specific evidence artifacts such as measurements, reports, quotes, or TA identity information. These artifacts are not used to move trust into the untrusted host. Rather, they provide the durable management and observability layer needed for multi-call OCI semantics, request binding, replay resistance, debugging, and evidence-aware verification. TEE-specific execution details are isolated behind a backend adapter, so the upper-level lifecycle and artifact layout remain stable across different TEE mechanisms.

We implement a Keystone-based prototype of \SYSNAME. Keystone is a suitable main backend because it exposes explicit protected-execution hooks and a clean host--TEE boundary, making it useful for studying OCI-compatible management of TEE-backed stages. In our prototype, \SYSNAME\ uses \texttt{crun} to run the REE-side anchor as an OCI-managed unit, while a host-side runtime service consumes stage requests, validates request metadata, allocates fresh EIDs, invokes Keystone \texttt{.ke} programs, and records responses and evidence artifacts. We evaluate the prototype using functional correctness tests, cold-start and end-to-end latency measurements, footprint and TCB analysis, and concurrent execution experiments. We further conduct case studies on SGX, TDX, and OP-TEE to examine whether the same lifecycle and stage abstraction can be mapped to enclave-style, VM-style, and embedded trusted-world TEEs.

This work makes the following contributions:
\begin{itemize}
  \item \textbf{A composite abstraction for OCI-managed confidential workloads.}
  We introduce the C4 abstraction, which models a confidential-computing workload as a composite of an REE-side anchor and one or more TEE-side confidential stages. This abstraction allows the standard OCI lifecycle to manage the composite as a single containerized execution instance, rather than exposing each protected invocation as an independent management object.

  \item \textbf{An OCI-compatible lifecycle and stage pipeline.}
  We design an OCI-compatible runtime structure that separates CID-level lifecycle management from EID-level confidential-stage execution. The runtime preserves multi-call OCI semantics while supporting on-demand TEE invocation through a stage pipeline that covers request claiming, preparation, backend execution, completion, failure handling, and response generation.

  \item \textbf{Durable artifacts for coordination, observability, and evidence binding.}
  We design a host-side artifact structure that records lifecycle state, request/response files, per-stage metadata, logs, and backend-specific evidence. These artifacts support deterministic lifecycle recovery, exactly-once stage processing, request binding, replay protection, and evidence-aware inspection without moving OCI lifecycle logic into the TEE.

  \item \textbf{A Keystone prototype and evaluation.}
  We implement \SYSNAME\ on Keystone and evaluate its functional correctness, startup and end-to-end overhead, footprint, and concurrent execution behavior. The results show that \SYSNAME\ introduces additional latency compared with native Keystone execution, but the overhead corresponds to explicit lifecycle mediation, request validation, EID allocation, backend dispatch, and artifact persistence, while the protected-side TCB remains narrow.

  \item \textbf{Portability case studies across heterogeneous TEEs.}
  We instantiate the same stage abstraction on SGX, TDX, and OP-TEE backends. These case studies show that \SYSNAME\ is not tied to Keystone: backend-specific details can be confined to adapters, while the upper-level lifecycle, request/response structure, and evidence-artifact model remain unchanged.
\end{itemize}

The rest of this paper is organized as follows. Section~\ref{sec:bg_assumption} introduces the background, assumptions, threat model, and scope. Section~\ref{sec:sys_design} presents the \SYSNAME\ system design, including the C4 abstraction, lifecycle model, and coordination rules. Section~\ref{sec:impl} describes the Keystone-based implementation and the key mechanisms used to preserve OCI-compatible behavior. Section~\ref{sec:sys_eval} evaluates functional correctness, runtime performance, footprint, TCB implications, and security implications. Section~\ref{sec:case-study} studies portability to SGX, TDX, and OP-TEE. Section~\ref{sec:related-work} discusses related work, and Section~\ref{sec:concl} concludes the paper.

\section{Background and Assumptions}\label{sec:bg_assumption}
\subsection{Container Execution Architecture}
Modern container platforms are organized around a stable interoperability boundary defined by the OCI Runtime Specification. A container instance is represented as an OCI bundle consisting of \texttt{config.json} (process and environment) and a \texttt{rootfs/} directory. The container manager invokes an OCI runtime via a fixed set of lifecycle operations (e.g., \texttt{create}, \texttt{start}, \texttt{kill}, \texttt{delete}, \texttt{state}, and \texttt{wait}) to materialize and control the workload.

Two properties of this boundary are critical to \SYSNAME. First, the OCI runtime interface is manager-agnostic: any runtime implementing the OCI contract can be plugged into existing container stacks without modifying the upper layers. Second, OCI lifecycle management is multi-call: repeated operations on the same instance are issued as separate runtime invocations, which requires the runtime to maintain minimal host-side state for status reporting and bookkeeping. \SYSNAME\ leverages this boundary to make C4 lifecycle management compatible with the standard container control flow, while keeping TEE-backed stage execution behind the runtime boundary.

\subsection{TEE-Backed Stage Model and Minimal Interface}
\SYSNAME\ targets TEEs that expose a protected execution context whose code and data remain isolated even if the host OS is compromised. For OCI integration, \SYSNAME\ does not require the OCI lifecycle to be implemented inside the TEE. It only requires a backend to provide a small set of hooks for preparing, entering or invoking, observing, and tearing down a protected stage execution. In Keystone, these hooks correspond to enclave primitives such as \emph{create}, \emph{load}, \emph{run}, and \emph{destroy}; in other TEEs, they may correspond to enclave entry, confidential-VM invocation, or TA command execution.

We deliberately adopt a minimal protected-side execution model. The TEE side contains only the runtime support and confidential-stage payload required by the backend, while OCI parsing, lifecycle coordination, bookkeeping, stream handling, and artifact management remain outside the protected context. This choice keeps the protected-side TCB narrow and allows TEE-backed stage execution to be managed through a C4 lifecycle rather than through a TEE-specific management plane.

\subsection{Assumptions, Threat Model, and Scope}
\label{sec:background-assumptions}

\paragraph{Assumptions}
We make the following assumptions. 
\textbf{(i) TEE correctness.} We assume that the underlying TEE correctly enforces the protected execution boundary for TEE-backed stages. This includes the hardware isolation mechanisms and the privileged TEE components that establish, enter, and maintain the protected context. In our Keystone prototype, this includes the security monitor and the minimal runtime support required by the protected-stage payload.
\textbf{(ii) Measurement and attestation availability.} We assume that the TEE platform can provide trustworthy measurement information for the protected execution context and, when required, expose an attestation interface to report such evidence to a verifier. \SYSNAME\ does not redesign the attestation protocol; instead, it preserves the lifecycle and artifact hooks needed to bind protected-stage execution to platform-provided evidence.
\textbf{(iii) Confidentiality and integrity focus.} We focus on adversaries that attempt to extract secrets, tamper with sensitive computation, or confuse the host-mediated stage invocation path. Denial-of-service is out of scope: a compromised host can always delay execution, drop requests, terminate processes, or refuse to schedule TEE-backed stages.
\textbf{(iv) Side channels out of scope.} We do not address microarchitectural or physical side channels, such as cache, speculative-execution, power, or electromagnetic leakage. These attacks require orthogonal mitigations from the underlying TEE platform or workload implementation.

\paragraph{Threat model}
We consider a strong software adversary that can compromise the host OS and all privileged software in the normal execution environment, including the container manager, OCI runtime services, host-side coordination logic, and ordinary host processes. The adversary may observe and modify host memory, file-system state, runtime metadata, request/response files, logs, and inter-process communication. It may also tamper with OCI bundles, manipulate host-side lifecycle records, reorder or replay host-side control operations, misroute stage requests, drop responses, or corrupt untrusted artifacts.

We assume that the adversary cannot break the TEE's hardware-enforced isolation or directly read or modify protected-stage memory and registers, except through the defined host--TEE interface. Therefore, the host-side state maintained by \SYSNAME, such as \texttt{state.json}, request/response files, \texttt{meta.json}, and \texttt{run.log}, is treated as untrusted management state rather than as a source of confidentiality or integrity by itself. Its role is to support OCI-compatible lifecycle continuity, observability, and evidence recording. Security-critical acceptance of a protected-stage request must instead rely on binding, freshness, ordering, and authentication checks, together with the native isolation and measurement mechanisms of the backend TEE.

\paragraph{Scope}
This paper focuses on enabling C4 composite workloads to be managed through the standard OCI runtime interface. \SYSNAME\ preserves the container manager's view of a workload as an OCI-managed unit, while internally separating CID-level composite lifecycle management from EID-level TEE-backed stage execution. Accordingly, our focus is on lifecycle compatibility, persistent coordination state, request/response mediation, stage-level artifacts, and backend-adapter integration.

\SYSNAME\ is designed to compose with, rather than replace, end-to-end confidential-computing mechanisms such as remote attestation, verifier policy, secret provisioning, and key management. These mechanisms can be bound to \SYSNAME's stage-level evidence artifacts, but their concrete protocols are outside the scope of this paper. Similarly, image integrity, rootfs protection, rollback prevention, and supply-chain verification are complementary to the runtime integration problem studied here.

Finally, \SYSNAME\ does not aim to reproduce the full Linux container feature set inside the TEE. The protected side remains narrow and contains only the backend-specific support and protected-stage payload needed for confidential computation, while OCI parsing, lifecycle bookkeeping, stream handling, logging, and artifact persistence remain outside the protected side.

\section{System Design}\label{sec:sys_design}

\subsection{\SYSNAME~Architecture}\label{subsec:ebcc-arch}

\begin{figure}[!t]
\centering
\includegraphics[width=3.5in]{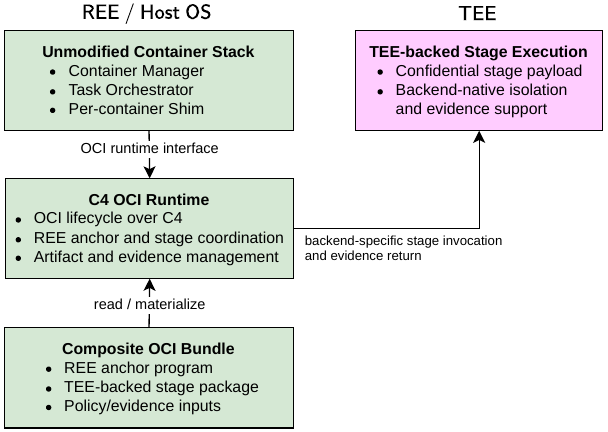}
\caption{Platform-independent architecture of \SYSNAME. \SYSNAME\ preserves the standard container stack and inserts a C4 OCI runtime at the OCI runtime boundary. The runtime materializes a composite OCI bundle, manages the C4 composite as one OCI-visible unit, and invokes TEE-backed confidential stages while receiving backend evidence.}
\label{fig:sys_arch_platindep}
\end{figure}

Figure~\ref{fig:sys_arch_platindep} presents the platform-independent architecture of \SYSNAME.
In typical confidential-computing applications, the REE-side program and the TEE-side confidential stages jointly form a single end-to-end workload: the REE side provides the surrounding business logic and external interactions, while the TEE side protects only the sensitive computation and data.
Managing only the protected execution component is therefore semantically incomplete.
Motivated by this necessity, we introduce the \emph{Containerized Confidential-Computing Composite (C4)} as the unit of deployment and management.

At a high level, \SYSNAME\ makes a C4 an OCI-managed execution unit.
The external OCI lifecycle (\texttt{create}/\texttt{start}/\texttt{state}/\texttt{wait}/\texttt{kill}/\texttt{delete}) operates on the composite as a whole, while the runtime internally coordinates the REE anchor, TEE-backed stage invocation, and artifact/evidence management.
This design preserves the standard container control plane and user-facing interfaces, while allowing sensitive computation to be executed as TEE-backed stages under a narrow protected-side interface.

As shown in Figure~\ref{fig:sys_arch_platindep}, \SYSNAME\ integrates with an unmodified container execution stack on the REE/host side.
The upper stack consists of the container manager, task orchestrator, and per-container shim.
These components continue to use the standard OCI runtime interface, and therefore do not need to be modified for C4 execution.
From their perspective, the \SYSNAME\ runtime behaves like a conventional OCI runtime: it consumes an OCI bundle and provides container-style lifecycle control, status query, waiting, and termination semantics.

The key architectural shift is that the OCI runtime manages a \emph{composite} rather than a single process or a single protected execution object.
The C4 OCI runtime preserves the OCI-visible lifecycle of the composite, coordinates the REE anchor and TEE-backed stage execution, and maintains artifact/evidence hooks needed for lifecycle continuity and later inspection.
Thus, OCI entry points are not treated as thin wrappers around a backend TEE lifecycle.
Instead, they are interpreted as composite-level management operations over the C4 instance.

The composite OCI bundle remains the packaging boundary between the container stack and the \SYSNAME\ runtime.
Besides conventional container configuration, the bundle carries the REE anchor program, the TEE-backed stage package, and policy/evidence inputs such as expected identity or serving requirements.
This avoids introducing a separate workload format for confidential stages: the standard bundle is still the materialization point, while \SYSNAME\ interprets its contents in a composite-aware manner.

On the TEE side, \SYSNAME\ assumes only that a backend can execute protected stages and provide backend-native isolation and evidence support.
The concrete backend may be an enclave-style TEE, a VM-style confidential execution environment, or an embedded trusted-world backend.
\SYSNAME\ does not require the full container runtime or general container-management logic to be placed inside the TEE.
Instead, sensitive computation is invoked as TEE-backed stage execution, while OCI parsing, lifecycle bookkeeping, artifact handling, stream/log compatibility, and orchestration-facing behavior remain on the REE side.

This separation is central to \SYSNAME's design.
The container stack continues to see a standard OCI-managed workload; the REE side retains the business lifecycle and compatibility logic; and the TEE side is used only for the protected computation that actually requires hardware-backed isolation.
As a result, \SYSNAME\ preserves operational compatibility with existing container tooling while keeping the protected-side trusted computing base narrow.

\subsection{Design Goals}\label{subsec:dsgn_goals}
Building on the C4 abstraction above, \SYSNAME\ aims to make composite confidential-computing workloads consumable through the standard container execution pipeline. The design is guided by four goals.

\paragraph{G1: OCI-level transparency}
C4 instances should be managed through the existing OCI runtime interface. The upper container manager stack (container manager, task orchestrator, and shim) remains unmodified, so \SYSNAME\ can be deployed as a drop-in runtime choice and adopted incrementally.

\paragraph{G2: Composite operational compatibility}
From the container manager's perspective, a C4 instance should behave like a conventional container task: it supports the standard lifecycle entry points (\texttt{create/start/kill/delete/state/wait}), provides stable multi-call behavior, reports well-formed status transitions and a single OCI-facing termination outcome, and exposes standard interaction and logging paths (e.g., \texttt{attach}/\texttt{logs}) so that existing operational tooling continues to work.

\paragraph{G3: Minimal TEE-side footprint}
The TEE side is intentionally small: only the backend runtime support and confidential-stage payload(s) required by the application are placed in the protected context. Container compatibility logic, such as OCI parsing, lifecycle coordination, bookkeeping, and stream plumbing, resides on the untrusted host side, keeping the protected-side TCB small and avoiding a heavyweight in-TEE container runtime.

\paragraph{G4: Backend portability across heterogeneous TEEs}
Although our prototype targets Keystone, \SYSNAME\ depends only on a small set of protected-stage execution hooks and a narrow host--TEE interface. This keeps the OCI-facing contract stable while allowing the backend adapter to be retargeted to enclave-style, VM-style, or embedded trusted-world TEEs with comparable invocation and evidence mechanisms.

\subsection{\SYSNAME~Runtime Structure}
\label{subsec:ebcc-runtime}

To make the discussion concrete, we instantiate \SYSNAME\ on Keystone and use it as the main prototype backend.
This choice is for exposition and prototype validation: Keystone provides explicit protected-execution primitives and a clear host--TEE boundary, while the runtime structure and OCI-facing contract remain independent of Keystone-specific mechanisms.
As long as a backend can provide protected-stage invocation hooks and a narrow host--TEE interaction path, the same C4 runtime structure can be retargeted to other TEEs.

\begin{figure}[!t]
\centering
\includegraphics[width=3.5in]{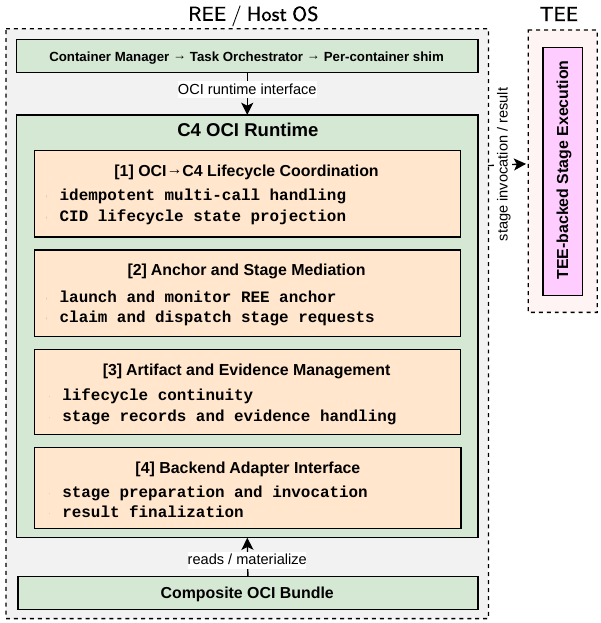}
\caption{Internal responsibilities of the C4 OCI runtime. The runtime maps OCI entry points to CID-level lifecycle coordination, launches and monitors the REE anchor, mediates stage requests, manages artifacts and evidence, and invokes TEE-backed stages through a backend adapter interface.}
\label{fig:keystone_oci_runtime_internals}
\end{figure}

\texttt{keystone-oci} is \SYSNAME's OCI-compatible runtime in the Keystone prototype.
Placed at the OCI runtime boundary, it connects an unmodified container stack to C4 execution without changing the container manager, task orchestrator, or per-container shim.
To the upper layers, \texttt{keystone-oci} behaves as a standard OCI runtime: it consumes a composite OCI bundle and implements the canonical entry points
(\texttt{create}, \texttt{start}, \texttt{state}, \texttt{wait}, \texttt{kill}, and \texttt{delete}).
Internally, however, the managed object is not a single process or a single protected execution context, but a C4 composite: the REE anchor carries the business lifecycle and external interaction, while TEE-backed stages are invoked on demand during the anchor's execution.

As shown in Figure~\ref{fig:keystone_oci_runtime_internals}, the runtime has four internal responsibilities.
First, \emph{OCI$\rightarrow$C4 lifecycle coordination} maps OCI entry points to CID-level lifecycle management.
The runtime interprets the bundle as a composite specification, materializes the persistent C4 state, and projects the internal CID state to OCI-visible states.
Because OCI operations are issued as separate runtime invocations, this layer must be idempotent and multi-call safe: repeated \texttt{create}, \texttt{start}, \texttt{state}, \texttt{wait}, \texttt{kill}, or \texttt{delete} calls must not duplicate side effects or produce inconsistent lifecycle results.

Second, \emph{anchor and stage mediation} coordinates the REE-side anchor and the TEE-backed stage path.
In \texttt{start}, the runtime launches or reattaches to the REE anchor as the primary execution driver.
During execution, the anchor emits stage requests when confidential computation is required.
The runtime claims these requests, checks their metadata, dispatches accepted requests to the backend execution path, and returns the corresponding result to the composite workflow.
This separates the OCI-visible lifecycle of the C4 instance from the intermittent protected-stage executions triggered by the anchor.

Third, \emph{artifact and evidence management} provides the continuity required by the OCI multi-call model and the observability required by stage-level execution.
The runtime maintains host-side artifacts that record lifecycle state, stage records, results, logs, and backend evidence.
These artifacts are operational and untrusted: they support lifecycle recovery, status reporting, debugging, and evidence inspection, but they are not themselves relied on to protect secrets or authenticate confidential computation.
Sensitive execution remains protected by the backend TEE, while the artifacts make the execution manageable through an OCI-compatible control path.

Fourth, the \emph{backend adapter interface} isolates TEE-specific execution details from the C4 runtime logic.
The runtime invokes protected stages through an adapter-style interface that prepares the stage context, performs the backend-specific protected invocation, and finalizes the result and evidence.
In the Keystone prototype, this interface is implemented using Keystone-backed \texttt{.ke} programs and the corresponding host--TEE execution path.
For other TEEs, the same runtime responsibilities can be preserved while the adapter maps stage preparation, invocation, and result finalization to backend-specific mechanisms such as enclave calls, confidential-VM requests, or TA commands.

This organization keeps container compatibility logic on the REE side and avoids embedding OCI lifecycle handling inside the TEE.
The C4 runtime provides the management layer required by the container stack, while the protected side only executes the confidential stages that require TEE isolation.
We discuss the security implications of host-side artifacts, multi-instance routing, and channel binding in Section~\ref{subsec:security-discussion}.

\subsection{System Model}\label{subsec:sys_model}

To reason about correctness and compatibility, we model EBCC as a composite execution unit rather than as a single process or a standalone enclave. As discussed in Section~\ref{subsec:ebcc-arch}, a C4 instance consists of an REE-side anchor program and one or more TEE-backed confidential stages invoked during the anchor's execution. This composite structure creates a semantic gap to OCI: the container stack exposes a process-oriented, multi-call lifecycle, whereas TEE-backed execution is stage-oriented and may occur intermittently within the REE program's timeline. The purpose of the model is to make this gap explicit: OCI operations act on the C4 composite, while protected computation is represented as separate stage invocations.

\begin{definition}[\textbf{C4 composite and stage instances}]
\label{def:c4-sm}
We model a C4 instance as
\[
C=\langle CID,\ \mathcal{R},\ \mathcal{T},\ r_{\textsf{anchor}}\rangle,
\]
where \(CID\in\Sigma^{*}\) is a globally unique composite identifier, \(\mathcal{R}\) and \(\mathcal{T}\) denote the REE-side and TEE-side component references, and \(r_{\textsf{anchor}}\in\mathcal{R}\) is the REE lifecycle anchor whose start and exit determine the C4 instance's OCI-visible execution boundary.

A C4 instance may invoke protected computation multiple times during the lifetime of the same anchor. We model each protected invocation as a stage instance identified by a fresh \(EID\). Let
\[
\mathcal{E}(C)=\{EID_1,EID_2,\ldots\}
\]
denote the set of stage instances created on demand for \(C\). Each \(EID\in\mathcal{E}(C)\) is associated with one protected-stage payload, one backend execution context, and one independent metadata/evidence record. This abstraction allows EBCC to reason about sequential and concurrent protected invocations while keeping the TEE-backed execution unit narrow.
\end{definition}

\begin{definition}[\textbf{CID-level lifecycle state}]
\label{def:cid-lifecycle}
The primary C4 lifecycle state space is
\[
S_{\textsf{all}}=\{\textsf{Init},\ \textsf{Prepared},\ \textsf{Running},\ \textsf{Stopped},\ \textsf{Failed}\}.
\]
\textsf{Init} denotes that no persistent C4 record exists. \textsf{Prepared} denotes that the instance has been materialized to the minimal persistent form required for multi-call OCI behavior. \textsf{Running} denotes that the REE anchor is alive. \textsf{Stopped} and \textsf{Failed} denote normal and failure termination, respectively.

The runtime maintains a persistent state record
\[
SR(CID):=\langle CID,\ S,\ ver\rangle,
\]
where \(S\in\{\textsf{Prepared},\textsf{Running},\textsf{Stopped},\textsf{Failed}\}\), and \(ver\in\mathbb{N}\) is a monotonically increasing version number used to order state updates. The absence of \(SR(CID)\) corresponds to \textsf{Init}. Entering \textsf{Running} requires that the lifecycle anchor has been established and is observable:
\[
SR(CID).S=\textsf{Running}\ \Rightarrow\ r_{\textsf{anchor}}\ \text{is defined and observable}.
\]

The runtime reports OCI-compatible states through the projection
\[
\pi_{\textsf{OCI}}(X)=
\begin{cases}
\textsf{created}, & X\in\{\textsf{Init},\textsf{Prepared}\},\\
\textsf{running}, & X=\textsf{Running},\\
\textsf{stopped}, & X\in\{\textsf{Stopped},\textsf{Failed}\}.
\end{cases}
\]
Auxiliary observability fields, such as trust status, health status, and TEE-stage phase, are reported separately and do not change the OCI projection. Additional details of these observability fields are given in Appendix~\ref{app:system-model-details}.
\end{definition}

\begin{definition}[\textbf{EID-level stage pipeline}]
\label{def:eid-stage-pipeline}
A protected stage invocation is modeled as an EID-level pipeline:
\[
\begin{aligned}
\textsf{ReqPending}
&\rightarrow \textsf{Claimed}
\rightarrow \textsf{Prepared} \\
&\rightarrow \textsf{Executing}
\rightarrow \textsf{Completed}/\textsf{Failed}.
\end{aligned}
\]
\textsf{ReqPending} means that the REE anchor has emitted a stage request. \textsf{Claimed} means that the runtime has atomically claimed the request, checked its metadata, and allocated a fresh \(EID\). \textsf{Prepared} means that the backend-specific protected execution context has been prepared and bound to the \(EID\). \textsf{Executing} means that the TEE-backed stage is running. \textsf{Completed}/\textsf{Failed} means that the runtime has recorded the result, metadata, logs, and response artifacts for the stage.

This stage pipeline is separate from the CID-level lifecycle. A C4 instance can remain \textsf{Running} while no protected stage is active, while one protected stage is executing, or while multiple \(EID\) instances are in flight. Conversely, a stage failure may cause the CID-level state to transition to \textsf{Failed} under the fail-fast policy.
\end{definition}

\begin{definition}[\textbf{OCI-to-C4 coordination}]
\label{def:coord-rules}
OCI entrypoints operate on the C4 composite identified by \(CID\), not directly on a backend TEE object. Each entrypoint reads the current \(SR(CID)\), performs the corresponding coordination action, and updates \(SR(CID)\) only through legal monotone state transitions.

The entrypoints have the following meanings. \texttt{create} materializes the persistent C4 state and initializes host-side artifacts, but does not execute a protected stage. \texttt{start} launches or reattaches to the REE anchor and moves the CID-level state to \textsf{Running}. \texttt{state} returns the OCI-projected state together with auxiliary observability fields. \texttt{wait} observes a terminal state and returns the composite exit result. \texttt{kill} terminates the anchor and cancels any ongoing protected stages. \texttt{delete} removes persistent artifacts after the instance reaches a terminal state.

All entrypoints are required to be idempotent with respect to the persisted state record: repeated invocations must not introduce illegal state transitions or duplicate side effects. Concurrent updates to the same \(CID\), and to each \(EID\) record when applicable, must be linearizable, for example through per-\(CID/EID\) mutual exclusion or version-checked updates.
\end{definition}

\begin{definition}[\textbf{Request binding invariant}]
\label{def:cb-inv}
Because stage requests are mediated by the untrusted host, every accepted request must be bound to the intended C4 instance and to the current session. For each \(CID\), there exist per-instance session parameters \((\mathit{epoch},sk)\) such that any accepted message \(m\) must satisfy
\[
\begin{aligned}
\mathsf{Accept}(m)\Rightarrow\ 
&\ \mathsf{Bind}(m,\ CID,\ \mathit{epoch})\\
&\wedge\ \mathsf{Auth}(m,sk)\\
&\wedge\ \mathsf{Fresh}(m,\mathit{epoch})\\
&\wedge\ \mathsf{Ordered}(m,\mathit{seq}).
\end{aligned}
\]
This invariant prevents a compromised host from misrouting, replaying, or reordering authenticated requests across C4 instances, except for denial-of-service behaviors. In the prototype, this invariant is instantiated using request identifiers, nonces, epochs, sequence numbers, response-path validation, and MAC verification.
\end{definition}

Detailed formal components that are not essential to the main design flow, including minimal state-record invariants, auxiliary observability semantics, detailed OCI entrypoint rules, and composite termination semantics, are provided in Appendix~\ref{app:system-model-details}.

In summary, the model separates three concerns that are otherwise easy to conflate. The CID-level lifecycle preserves OCI-compatible management of the composite instance. The EID-level stage pipeline captures on-demand TEE-backed execution. The request binding invariant constrains the host-mediated interaction path so that stage execution remains tied to the intended C4 context.

\section{Implementation}\label{sec:impl}

We implement a prototype of \SYSNAME\ on top of Keystone, a representative open-source enclave framework in the RISC-V ecosystem.
Keystone provides an explicit security monitor (SM) and a small set of protected-execution primitives, including \emph{create}, \emph{load}, \emph{run}, and \emph{destroy}.
This makes it suitable for implementing and instrumenting the boundary between an untrusted host-side runtime and protected Keystone enclave instances.

The prototype instantiates the C4 composite abstraction described in Section~\ref{sec:sys_design}.
The OCI-managed object is the composite identifier \(CID\).
For each \(CID\), the REE-side anchor process \(r_{\textsf{anchor}}\) carries the externally visible business lifecycle, while confidential computation is issued as on-demand stage requests.
Each accepted stage request is mapped to a fresh EID-level Keystone execution instance.
Thus, the implementation separates two levels of management: CID-level lifecycle handling for OCI compatibility, and EID-level protected execution for confidential stages.

Although this section focuses on Keystone, the implementation follows the backend-adapter principle of \SYSNAME.
The host-side OCI integration does not depend on Keystone-specific semantics.
It requires only that a backend can prepare a protected stage context, invoke the stage, return execution results, and provide backend evidence or metadata.
In the Keystone prototype, these operations are realized through Keystone enclave instances and \texttt{.ke} stage payloads.

\subsection{Keystone-based EBCC Architecture}
\label{subsec:impl-arch}

\begin{figure*}[!t]
\centering
\includegraphics[width=5in]{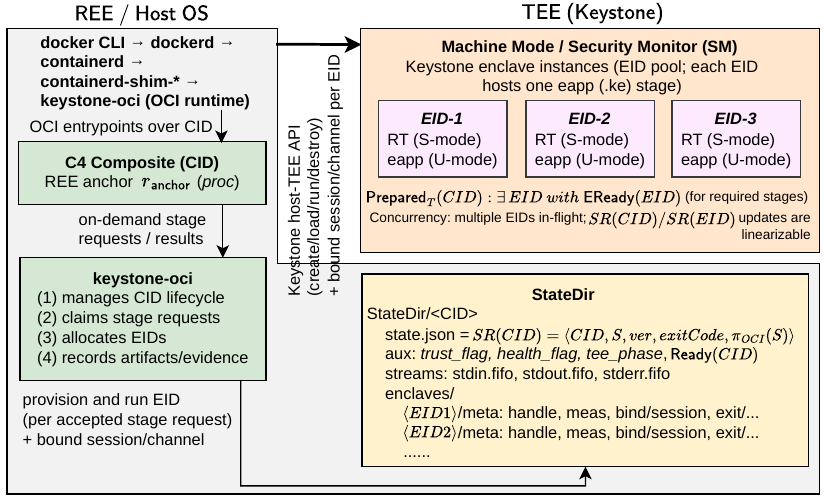}
\caption{Keystone-based implementation of \SYSNAME. The \texttt{keystone-oci} runtime manages the CID-level C4 lifecycle on the REE side, claims stage requests from the REE anchor, allocates EID-level Keystone enclave instances, and records CID- and EID-level artifacts in \texttt{StateDir}.}
\label{fig:sys_arch_imp_keystone}
\end{figure*}

Figure~\ref{fig:sys_arch_imp_keystone} shows the Keystone-based implementation.
The upper-left part follows the standard container execution path: \texttt{docker}, \texttt{containerd}, and the per-container shim invoke \texttt{keystone-oci} as an OCI runtime.
The upper container stack is not modified.
All standard OCI entrypoints are still issued over the managed \(CID\), so the container stack observes a C4 composite as an ordinary OCI-managed unit.

Inside the C4 composite, the REE anchor process \(r_{\textsf{anchor}}\) is the primary lifecycle anchor.
It is launched and monitored by \texttt{keystone-oci}, and it emits stage requests when confidential computation is required.
These requests are not exposed as independent container objects.
Instead, \texttt{keystone-oci} claims them, checks the associated request/session context, allocates EIDs, invokes the corresponding Keystone stage, and records the result back into the same \(CID\)-scoped execution context.

On the TEE side, Keystone executes confidential stages as enclave instances managed by the security monitor.
Each EID hosts one \texttt{.ke} stage payload, together with the Keystone runtime support required to execute that payload.
Multiple EIDs may be in flight when the anchor issues concurrent stage requests.
This directly realizes the model-level separation between the CID-level lifecycle and the EID-level stage pipeline: the \(CID\) remains the OCI-visible management object, while each \(EID\) represents one protected stage invocation.

The bottom-right part of Figure~\ref{fig:sys_arch_imp_keystone} shows the persistent \texttt{StateDir}.
This directory is the implementation substrate for multi-call OCI semantics and stage-level artifact persistence.
At the CID level, \texttt{state.json} records the persistent state record
\[
SR(CID)=\langle CID,S,ver,exitCode,\pi_{\textsf{OCI}}(S)\rangle,
\]
together with auxiliary observability fields such as trust state, health state, TEE phase, and readiness.
At the stage level, \texttt{StateDir} contains per-EID metadata, logs, binding/session information, backend measurement fields, and exit observations.
In this way, the implementation preserves a durable link among the OCI-visible composite instance, the REE anchor, each protected stage execution, and the evidence artifacts generated during execution.

\subsection{EBCC Runtime Implementation}
\label{subsec:runtime-impl}

The main implementation logic resides in \texttt{keystone-oci}.
Rather than treating a Keystone enclave as the stable OCI-managed object, \texttt{keystone-oci} treats the C4 composite as the managed object and uses \texttt{StateDir/<CID>} as the persistent coordination context.
This context allows separate OCI invocations to resolve to the same composite instance and observe consistent lifecycle state, exit status, and stage records.

The \texttt{create} path materializes the C4 instance.
It initializes the CID-level state record, prepares the request/response path, creates the artifact directories, records the bundle association, and initializes the session or binding state needed for later stage requests.
It does not create a long-lived Keystone enclave instance.
This distinction is important: \texttt{create} prepares the composite for OCI management, while protected execution is deferred until the anchor emits a stage request.

The \texttt{start} path launches or reattaches to the REE anchor process.
Once the anchor becomes observable, the \(CID\)-level state is advanced to the running state and the composite becomes visible to the upper container stack as an executing workload.
During this period, the TEE side may still be idle.
A protected stage is entered only when the anchor produces an on-demand stage request.

When a stage request appears, \texttt{keystone-oci} performs the stage-processing path.
It claims the request, validates the request/session metadata, allocates a fresh EID, and invokes the corresponding Keystone \texttt{.ke} stage payload through the Keystone host--TEE API.
The protected execution is then provisioned and run under a bound session or channel associated with the accepted request.
After the stage returns, the runtime records the result, exit information, backend metadata, logs, and evidence-related fields under the corresponding EID artifact record.
The response is then made available to the REE anchor through the request/response path.

The remaining OCI entrypoints are resolved against the same persistent CID-level context.
The \texttt{state} path reports the OCI projection of the current \(CID\)-level state, together with auxiliary observability fields when available.
The \texttt{wait} path observes a stable terminal state and returns the recorded composite exit result.
The \texttt{kill} path terminates the anchor and cancels any outstanding stage executions when applicable.
The \texttt{delete} path removes the persistent coordination context after the instance has reached a valid terminal state.
Across these entrypoints, \texttt{keystone-oci} maintains monotone and idempotent lifecycle behavior so that repeated or retried OCI invocations do not duplicate stage effects or corrupt the recorded lifecycle state.

This implementation structure also determines how concurrency is handled.
Concurrent protected computation is represented as multiple EIDs associated with the same CID.
Each EID has its own stage metadata, backend handle, binding/session context, log, and exit observation.
The CID-level state and EID-level records are updated through linearizable bookkeeping so that the composite remains stable from the OCI perspective even when multiple stage requests are in flight.

\subsection{StateDir and Evidence Artifacts}
\label{subsec:impl-artifacts}

The \texttt{StateDir} layout is central to the implementation.
It is not a trusted security boundary by itself; rather, it is the persistent host-side substrate that makes EBCC compatible with the multi-call OCI runtime model.
Because OCI operations are invoked as separate processes, runtime continuity cannot depend on transient process memory.
The runtime must recover the managed object, its lifecycle state, and its most recent execution outcome from persistent artifacts.

At the CID level, \texttt{StateDir/<CID>} stores the lifecycle state, version number, exit code, and OCI-projected status.
These fields allow later \texttt{state}, \texttt{wait}, \texttt{kill}, and \texttt{delete} operations to refer to the same composite instance.
Auxiliary fields record additional information such as trust status, health status, TEE phase, and readiness.
These auxiliary fields refine observability but do not replace the primary OCI lifecycle state.

At the EID level, the runtime records per-stage artifacts under the corresponding enclave/stage directory.
Each accepted stage request receives a fresh EID and produces metadata describing the selected stage, backend handle, binding or session information, measurement-related fields, return code, and exit status.
Execution output is recorded as a run log, and the stage result is returned through the response path.
These artifacts provide a durable audit trail from a host-visible stage request to a specific TEE-backed execution and its result.

The request/response path and the per-EID artifacts together provide the concrete implementation of stage-level evidence binding.
They allow EBCC to associate each protected execution with the current C4 instance, the accepted request, and the generated backend evidence.
However, these artifacts remain in the untrusted host environment.
They support lifecycle continuity, debugging, evidence inspection, and replay detection, but the confidentiality and integrity of sensitive computation still rely on the Keystone protected execution boundary and the corresponding request/session binding checks.

\subsection{Key Mechanisms and Challenges}

\subsubsection{Lifecycle Mapping and Semantic Consistency}

A central challenge is that OCI lifecycle operations are defined over a stable managed object, whereas protected computation in EBCC is realized through intermittent EID-level TEE-backed stage invocations.
\SYSNAME\ resolves this mismatch by preserving lifecycle identity at the C4 composite level.
The OCI entrypoints \texttt{create}, \texttt{start}, \texttt{state}, \texttt{wait}, \texttt{kill}, and \texttt{delete} all operate on the same \(CID\)-scoped composite instance.
Stage completion, rejection, or failure is reflected back into this instance through the persistent state and artifact records, rather than being exposed as the lifecycle of a separate backend object.
This is the key condition for keeping OCI-visible behavior stable despite on-demand TEE-backed execution.

\subsubsection{Persistent State Across OCI Calls}

A second challenge comes from the multi-call nature of the OCI runtime interface.
Because each lifecycle operation is issued as a separate runtime invocation, OCI-visible continuity cannot be derived from transient backend state alone.
The implementation therefore uses \texttt{StateDir/<CID>} to preserve CID-level lifecycle state, request/response records, and EID-level artifacts across calls.
Externally visible state transitions are reported only after the corresponding lifecycle or stage event has been durably recorded.
This prevents later OCI calls from observing a state that has not been materialized in the coordination context.

\subsubsection{Stage Interaction, Observability, and Request Binding}

A third challenge is the host-mediated interaction path between the REE anchor and TEE-backed stage execution.
The REE anchor emits stage requests in the untrusted environment, and \texttt{keystone-oci} dispatches accepted requests to Keystone EID instances.
This separation is necessary for preserving OCI compatibility, but it also creates possible semantic attacks: a compromised host may replay stale requests, misroute a request, bind a response to the wrong CID/EID, or drop a request entirely.

To address this, EBCC treats stage acceptance as an explicit validation step.
Each accepted request is bound to the current C4 context and to a session or channel associated with the allocated EID.
The runtime validates request/session metadata before dispatching protected execution and records the resulting metadata and evidence under the corresponding EID artifact path.
This does not prevent denial-of-service by a compromised host, but it prevents stale, malformed, or misrouted requests from being silently accepted as legitimate stage executions for the wrong composite context.

\subsubsection{Protected-side Minimality}

The implementation deliberately avoids moving OCI lifecycle handling, bundle parsing, stream plumbing, or artifact management into the protected side.
These functions remain on the REE side in \texttt{keystone-oci} and \texttt{StateDir}.
The Keystone protected context contains only the runtime support and stage payload needed to execute the confidential computation.
This keeps the protected-side TCB narrow and ensures that container compatibility is achieved through host-side management rather than by embedding a container runtime inside the TEE.

\section{Evaluation}\label{sec:sys_eval}
\subsection{Experimental Setup}
\label{subsec:experimental-setup}

\textbf{Platform and execution model.}
All main experiments are conducted on an Intel TDX-capable server equipped with an Intel Xeon Platinum 8558 processor (48 cores @ 2.1\,GHz), 256\,GB DDR5-4800 memory, and a 2\,TB Samsung SSD 990 EVO Plus, running Ubuntu 24.04. On this physical host, we run a Keystone-enabled RISC-V guest environment in QEMU to evaluate the Keystone backend. The main Keystone experiments therefore use a QEMU-based RISC-V execution environment on a TDX-capable server; TDX itself is evaluated separately as one of the heterogeneous TEE backends in the case studies.

EBCC uses an OCI runtime, \texttt{crun}, to run the C4 anchor as an OCI-managed unit inside the guest. The anchor is a resident process that emits stage requests according to the configured workload. A host-side EBCC runtime service, running in the same guest, consumes these requests, validates them, allocates EIDs, maps stage names to Keystone \texttt{.ke} programs, invokes the corresponding Keystone backend execution, and then writes the response and evidence artifacts. This setup allows the OCI-visible lifecycle to be managed independently from the TEE-specific execution path, while still binding each confidential stage execution to durable artifacts.

\textbf{Workloads.}
We use two representative Keystone stage workloads. The light workload is \texttt{hello.ke}, a \texttt{hello world} program with a small amount of additional computation. It reflects the basic cost of invoking a TEE-backed stage through the EBCC-managed path. The heavy workload is \texttt{aesgcm16m.ke}, which performs AES-128-GCM encryption over 16\,MiB of data inside the Keystone execution context. It represents a longer confidential computation and is used to examine how EBCC's fixed management cost behaves when the protected workload itself becomes more substantial.

In EBCC, a \emph{stage} denotes one managed confidential task invocation. For each stage request, the anchor emits a request containing fields such as the stage name, epoch, sequence number, request ID, nonce, response path, and MAC. The EBCC service loop claims the request, checks the request metadata, prevents replay using request IDs and nonces, verifies the epoch/sequence/MAC fields, allocates a fresh EID, executes the mapped Keystone \texttt{.ke} program, and records the result. Therefore, in the concurrency experiment, the concurrency level $k$ means that $k$ EBCC-managed confidential stages are issued and completed in one experimental round.

\textbf{Measurements and reporting.}
For cold-start latency, we measure the \textit{create}, \textit{start}, and combined \textit{bringup} latency. The \textit{create} phase includes the preparation of per-CID runtime state, request/response directories, enclave-artifact directories, lock files, EID sequence state, OCI bundle, rootfs copy, anchor configuration, session information, and the initial \texttt{state.json}. The \textit{start} phase launches the OCI-managed anchor process through \texttt{crun}. For end-to-end latency, we measure the full completion time from issuing a managed stage request to receiving the corresponding response after backend execution and artifact persistence. For concurrent execution, we report both the total elapsed time of each round and the effective stage throughput, computed as $k/\textit{elapsed\_s}$ with the unit of stages/s.

Unless otherwise stated, latency experiments are repeated 50 times and reported using median, P95, mean, and standard deviation. The concurrency experiment is repeated over multiple rounds for each concurrency level, and we report median, P95, and mean values for both elapsed time and effective stage throughput.

\textbf{Evidence-artifact-driven validation.}
EBCC materializes machine-readable artifacts under \texttt{StateDir/<CID>/}, including \texttt{state.json}, \texttt{session.json}, \texttt{requests/}, \texttt{responses/}, \texttt{enclaves/<EID>/meta.json}, \texttt{enclaves/<EID>/run.log}, and \texttt{anchor.out}. Functional correctness is validated by checking these artifacts rather than relying on screenshots or manual log inspection. In particular, successful execution requires that each accepted stage request is bound to a fresh EID, produces a response, records the backend output in \texttt{run.log}, updates per-stage metadata in \texttt{meta.json}, and leaves the CID-level state consistent with the OCI-managed lifecycle.

\subsection{Functional Correctness Evaluation}
\label{subsec:functional-correctness}

We evaluate functional correctness by checking whether EBCC (i) faithfully implements the C4 instance lifecycle semantics, (ii) correctly closes the request--response loop for on-demand confidential stages, and (iii) produces complete and consistent evidence artifacts that support mechanical validation. Each run materializes a \texttt{StateDir/<CID>/} directory containing \texttt{state.json}, request/response files, per-stage metadata (\texttt{meta.json}), execution traces (\texttt{run.log}), and anchor outputs. Table~\ref{tab:functional_correctness_evaluation} summarizes the results: each row corresponds to a correctness test (lifecycle workflow, artifact/invariant checks, state-field consistency, and concurrent serving), while the columns report the number of rounds and the associated metrics, including workflow completion rate (WCR), command success rate (CSR), state consistency rate (SCR), and invariant pass rate (IPR), with notes indicating the workload structure (e.g., four stages per round) and what each check covers.

\paragraph{Lifecycle workflow correctness}
The first row in Table~\ref{tab:functional_correctness_evaluation} reports the correctness of the end-to-end lifecycle workflow over 100 rounds. In each round, EBCC executes a complete \texttt{create$\rightarrow$start$\rightarrow$serve$\rightarrow$wait$\rightarrow$kill$\rightarrow$delete} cycle and processes four stage requests. We observe a \textbf{100\%} workflow completion rate (WCR) and a \textbf{100\%} command success rate (CSR), showing that the lifecycle operations are correctly implemented and that the OCI-managed anchor can be repeatedly created, started, terminated, and deleted without leaving stale state.

\paragraph{Artifact completeness and consistency}
The second row of Table~\ref{tab:functional_correctness_evaluation} evaluates whether EBCC consistently generates the expected evidence artifacts and whether the associated consistency checks pass. Across 100 rounds, the invariant pass rate (IPR) is \textbf{100\%}. For each stage execution, EBCC creates a fresh EID directory and records a well-formed \texttt{meta.json} and the corresponding \texttt{run.log}. Each request yields exactly one response, and the recorded per-stage metadata (e.g., \texttt{stage}, \texttt{example}, \texttt{rc}, and request binding) matches the execution outcome. This indicates that EBCC maintains complete and internally consistent per-stage records across repeated runs.

\paragraph{State-field consistency}
The third row of Table~\ref{tab:functional_correctness_evaluation} evaluates whether \texttt{state.json} remains consistent with the observed execution results. Over 100 trials, the state consistency rate (SCR) is \textbf{100\%}. Specifically, the recorded state transitions and summary fields, such as terminal status, last executed stage, last return code, and last EID, are consistent with the corresponding runtime events, including anchor termination and response generation. These results indicate that \texttt{state.json} provides a reliable summary of the execution state across repeated runs.

\paragraph{Concurrent request serving}
The last row of Table~\ref{tab:functional_correctness_evaluation} evaluates EBCC under concurrent runtime services. We run 100 rounds with multiple \texttt{serve} instances operating on the same CID, and obtain an IPR of \textbf{100\%}, with all requests consumed \emph{exactly once}. This shows that EBCC preserves correct request claiming and EID allocation under concurrency: concurrent \texttt{serve} loops neither duplicate stage execution nor overwrite per-stage artifacts, and they do not produce conflicting outcomes for the same request.

\paragraph{Summary}
Overall, Table~\ref{tab:functional_correctness_evaluation} shows that EBCC remains correct under both sequential and concurrent execution. Across all reported tests, lifecycle operations complete successfully, evidence artifacts remain complete and consistent, \texttt{state.json} stays aligned with the observed execution results, and concurrent \texttt{serve} instances preserve exactly-once request consumption.

\newcolumntype{Y}{>{\raggedright\arraybackslash}X}
\begin{table*}[t]
\centering
\caption{Functional Correctness Evaluation}
\label{tab:functional_correctness_evaluation}
\begin{tabularx}{\textwidth}{%
>{\hsize=0.75\hsize\raggedright\arraybackslash}X
c c c c c
>{\hsize=1.25\hsize\raggedright\arraybackslash}X}
\toprule
\textbf{Test} & \textbf{Rounds} & \textbf{WCR} & \textbf{CSR} & \textbf{SCR} & \textbf{IPR} & \textbf{Notes} \\
\midrule
Lifecycle workflow          & 100 & \textbf{100\%} & \textbf{100\%} & --             & --             & 4 stages per round \\
Artifact / invariant checks & 100 & \textbf{100\%} & --             & --             & \textbf{100\%} & \texttt{meta}/\texttt{log}/\texttt{rc} are consistent and complete \\
State-field consistency     & 100 & --             & --             & \textbf{100\%} & --             & \texttt{state.json} is consistent with runtime behavior \\
Concurrent serve            & 100 & \textbf{100\%} & --             & --             & \textbf{100\%} & Requests are consumed exactly once \\
\bottomrule
\end{tabularx}
\end{table*}

\subsection{Performance Evaluation}
\label{subsec:performance}

We next evaluate the runtime performance overhead of EBCC in the current prototype. Because EBCC adds OCI-compatible lifecycle coordination around confidential-stage execution, its cost may appear in different parts of the execution path. We therefore focus on three runtime aspects: cold-start latency, end-to-end completion latency, and concurrent execution behavior. Resource footprint and TCB implications are discussed separately in Section~\ref{subsec:footprint-tcb}.

\subsubsection{Cold-Start Latency}

\begin{figure}[!t]
\centering
\captionsetup[subfloat]{font=scriptsize,skip=1pt}

\subfloat[Light workload.\label{fig:cold-start-light}]{
    \includegraphics[width=0.96\columnwidth]{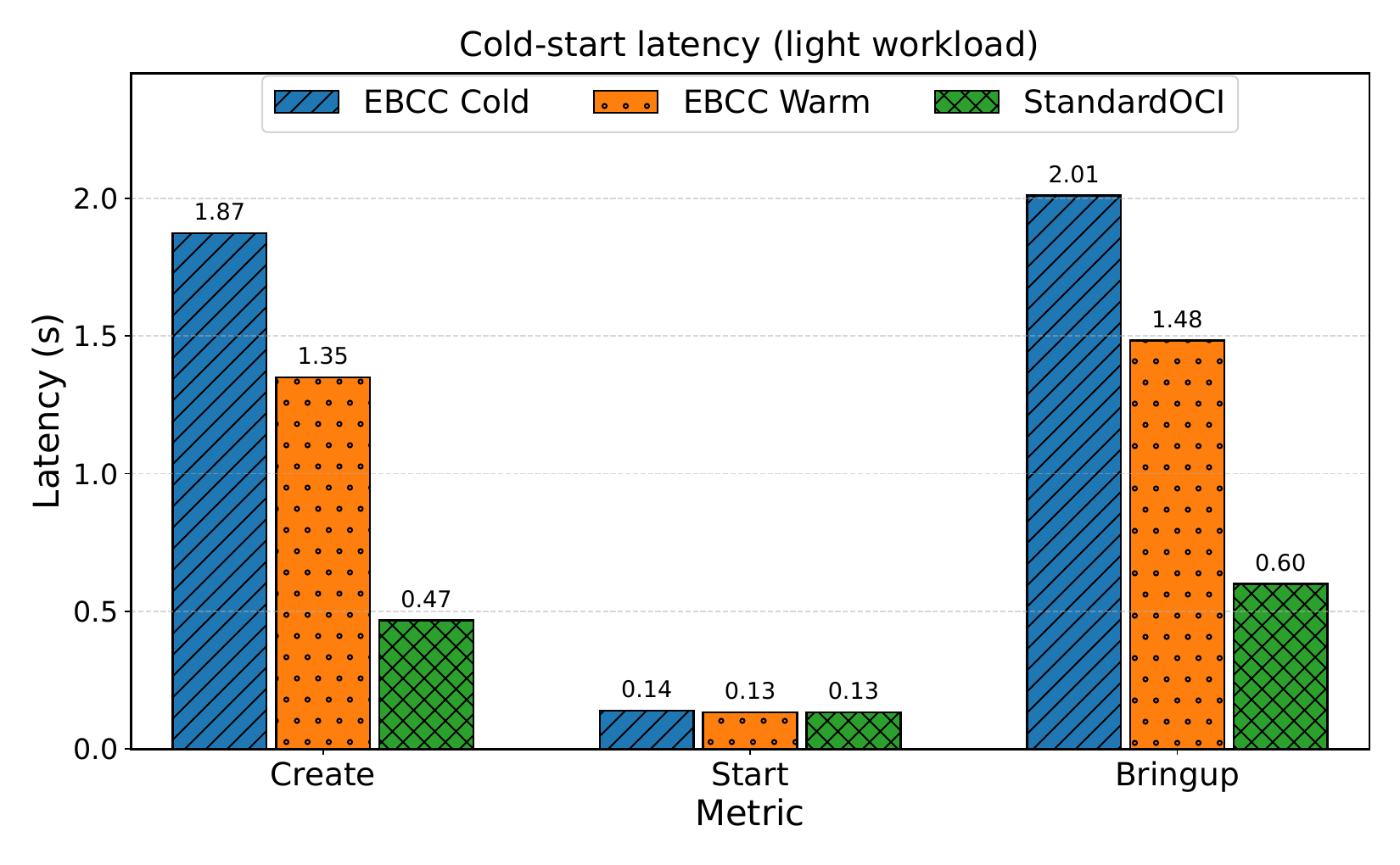}
}

\vspace{-0.8em}

\subfloat[Heavy workload.\label{fig:cold-start-heavy}]{
    \includegraphics[width=0.96\columnwidth]{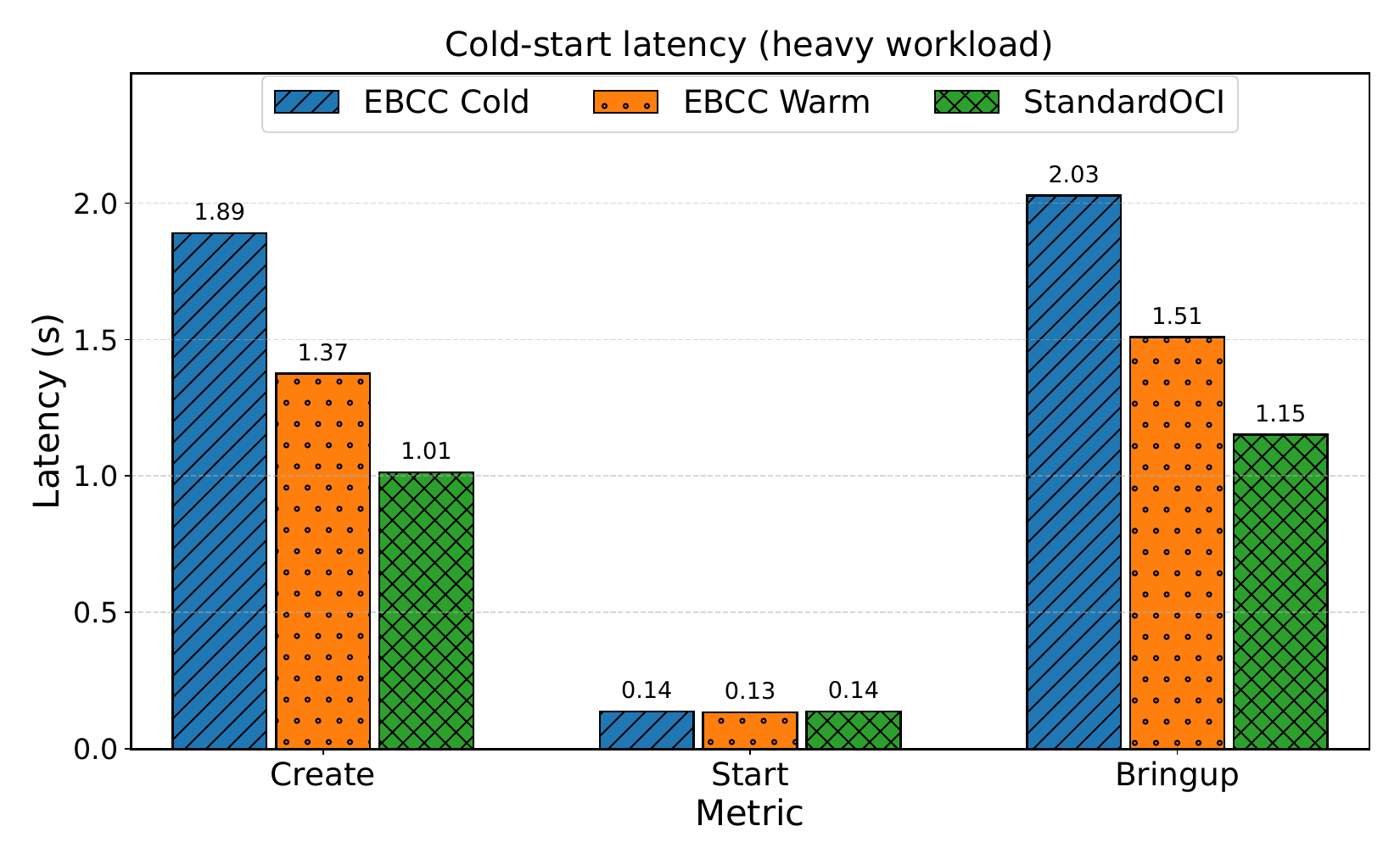}
}

\caption{Cold-start latency breakdown under light and heavy workloads.}
\label{fig:cold-start-latency}
\end{figure}

Figure~\ref{fig:cold-start-latency} breaks down the cold-start path into \textit{create}, \textit{start}, and the combined \textit{bringup} latency. In the EBCC prototype, \textit{create} is not merely a container-bundle creation step. It also prepares the per-CID runtime state, including the request and response directories, enclave-artifact directory, lock files, EID sequence file, OCI bundle, rootfs copy, anchor configuration, initial \texttt{state.json}, and session information. By contrast, \textit{start} mainly launches the OCI anchor process through the OCI runtime and moves the instance into the running state. This implementation structure explains why the additional cold-start cost is concentrated in \textit{create}, while \textit{start} remains small across configurations.

Under the light workload, \texttt{EBCC Cold} has a median \textit{bringup} latency of about 2.01~s, while \texttt{EBCC Warm} reduces it to about 1.48~s. The reduction mainly comes from the \textit{create} phase: the median \textit{create} latency decreases from about 1.87~s in \texttt{EBCC Cold} to about 1.35~s in \texttt{EBCC Warm}. In contrast, the median \textit{start} latency remains close across \texttt{EBCC Cold}, \texttt{EBCC Warm}, and \texttt{StandardOCI}, all around 0.13--0.14~s. This shows that warm execution mainly avoids part of the repeated state and bundle preparation cost, but it does not significantly change the final OCI start operation.

The heavy workload shows the same structure. The median \textit{bringup} latency decreases from about 2.03~s in \texttt{EBCC Cold} to about 1.51~s in \texttt{EBCC Warm}, and the median \textit{create} latency drops from about 1.89~s to about 1.37~s. The \textit{start} latency again remains around 0.13--0.14~s. Compared with \texttt{StandardOCI}, EBCC still has a clear \textit{bringup} overhead: \texttt{StandardOCI} takes about 0.60~s under the light workload and about 1.15~s under the heavy workload. The gap is expected because EBCC maintains additional per-instance coordination state and prepares the execution context required for later confidential-stage invocation, rather than only starting a conventional OCI container.

Overall, the cold-start results show that EBCC's startup overhead is dominated by the preparation of an OCI-managed confidential execution instance. Warm execution reduces the repeated \textit{create} cost, but the remaining gap between \texttt{EBCC Warm} and \texttt{StandardOCI} reflects EBCC's structural requirements: persistent CID-level state, request/response paths, EID tracking, session state, and artifact directories must be established so that later TEE-backed stages can be invoked, recorded, and bound to the OCI-visible lifecycle.

\subsubsection{End-to-End Completion Latency}

\begin{figure}[!t]
\centering
\captionsetup[subfloat]{font=scriptsize,skip=1pt}

\subfloat[Light workload.\label{fig:e2e-light}]{
    \includegraphics[width=0.96\columnwidth]{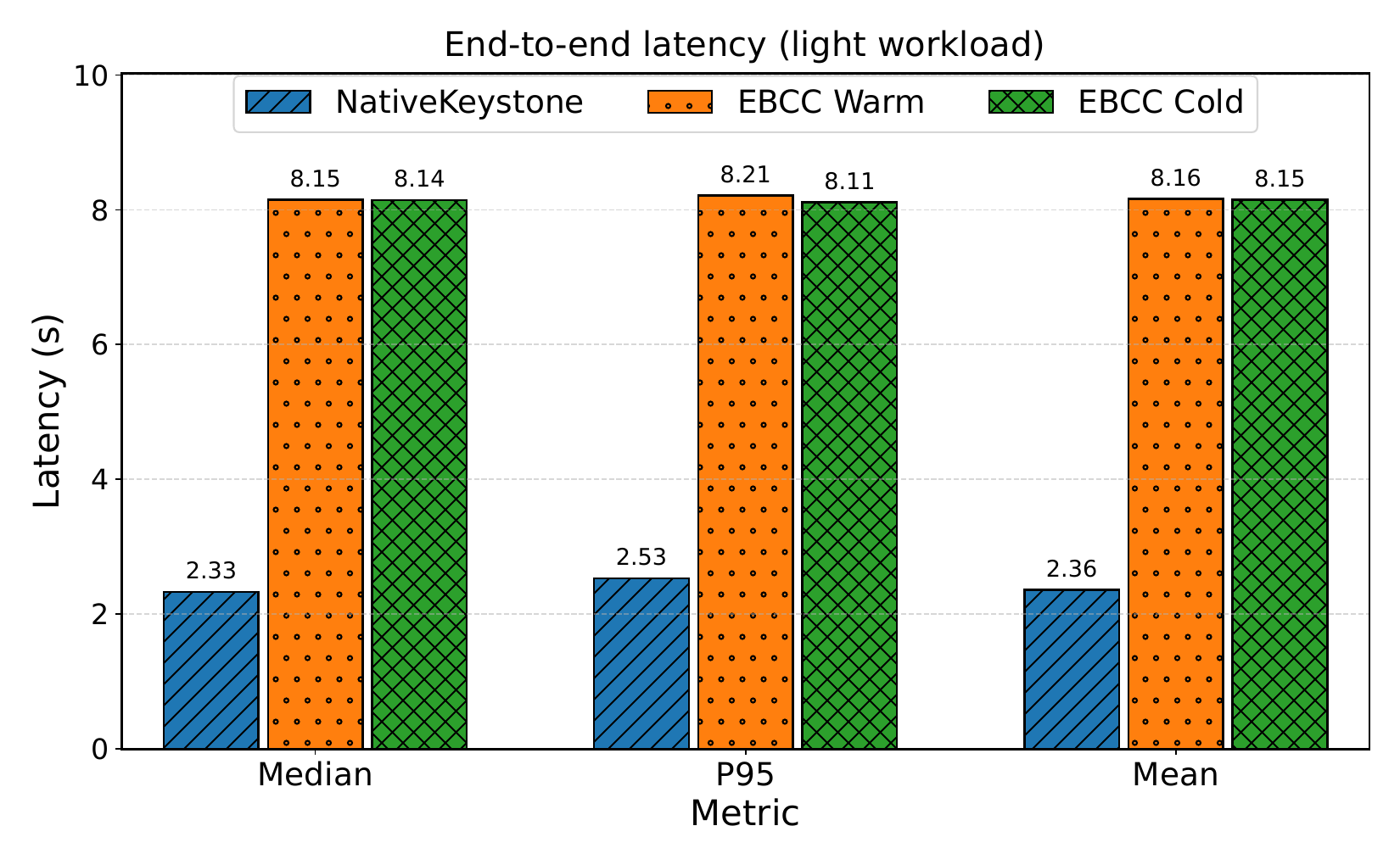}
}

\vspace{-0.8em}

\subfloat[Heavy workload.\label{fig:e2e-heavy}]{
    \includegraphics[width=0.96\columnwidth]{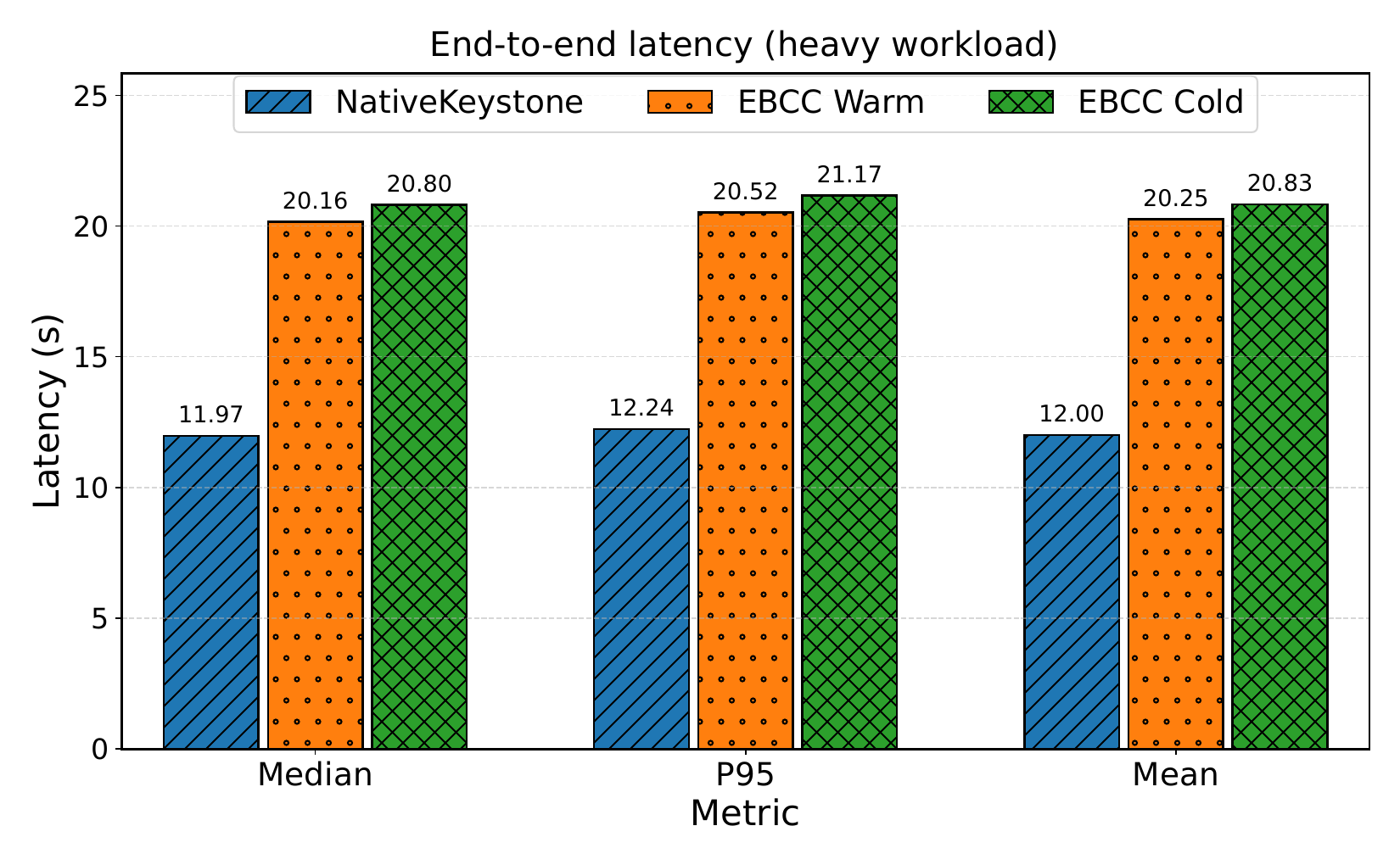}
}

\caption{End-to-end completion latency under light and heavy workloads.}
\label{fig:e2e-overhead}
\end{figure}

Figure~\ref{fig:e2e-overhead} reports the end-to-end completion latency of native Keystone and EBCC-managed execution. Unlike the cold-start measurement, the end-to-end path includes not only instance bring-up, but also the complete managed stage-processing path. In the prototype, an anchor emits a stage request with an epoch, sequence number, request ID, nonce, response path, and MAC. The EBCC \texttt{serve}/\texttt{poll} loop then claims the request, validates the response path, checks replay state, verifies the epoch, sequence number, and MAC, allocates a fresh EID, maps the stage name to a Keystone example, executes the backend program, records \texttt{run.log} and \texttt{meta.json}, updates \texttt{state.json}, advances the session sequence, and finally writes the response. Therefore, the end-to-end latency captures the full cost of making a Keystone execution stage observable and manageable through the EBCC lifecycle.

Under the light workload, native Keystone completes in about 2.33~s at the median, while \texttt{EBCC Cold} and \texttt{EBCC Warm} both take about 8.14--8.15~s. The cold--warm difference is negligible in this setting. This indicates that the dominant overhead is not the cold-start component alone. Instead, most of the gap comes from the persistent managed path around each stage execution, including request claiming, validation, EID allocation, state transitions, artifact generation, and response synchronization between the anchor and the EBCC service loop. For a short \texttt{hello}-style workload, this fixed management path is large relative to the protected computation itself, so the EBCC overhead is especially visible.

Under the heavy workload, native Keystone has a median completion latency of about 11.97~s. \texttt{EBCC Cold} increases the median latency to about 20.80~s, while \texttt{EBCC Warm} reduces it to about 20.16~s. Warm execution therefore saves about 0.64~s at the median and also reduces the P95 latency from about 21.17~s to about 20.52~s. This reduction is consistent with the cold-start breakdown: warm execution mainly removes part of the repeated preparation cost. However, the remaining gap between \texttt{EBCC Warm} and native Keystone is still much larger than the cold--warm difference. This shows that the main end-to-end cost is the structural cost of executing Keystone stages through EBCC's OCI-compatible coordination and artifact-preserving path.

The comparison between the light and heavy workloads further clarifies the role of workload size. For the light workload, the EBCC management path dominates the total latency because the backend computation is short. For the heavy workload, the native Keystone execution time increases substantially, so part of the fixed EBCC cost is amortized by the longer protected computation. Nevertheless, EBCC still remains slower than native Keystone because every stage must pass through the same controlled path: request authentication, replay protection, EID assignment, backend invocation, metadata persistence, state update, and response generation. These operations are the price of making a TEE-backed execution stage compatible with an OCI-managed lifecycle.

Overall, the end-to-end results show two layers of overhead. The first is the cold-start component, which can be reduced by warm execution, especially under the heavy workload. The second is the persistent warm-path overhead caused by EBCC's stage-level coordination and evidence-preserving execution path. This overhead is not an accidental implementation artifact; it corresponds to the design goal of binding confidential-stage execution to durable lifecycle state, per-stage artifacts, and controlled host--TEE interaction. EBCC therefore trades additional latency for a managed execution structure that native Keystone does not provide.

\subsubsection{Concurrent Execution Performance}

\begin{figure}[!t]
\centering
\captionsetup[subfloat]{font=scriptsize,skip=1pt}

\subfloat[Total elapsed time.\label{fig:concurrent-elapsed}]{
    \includegraphics[width=0.96\columnwidth]{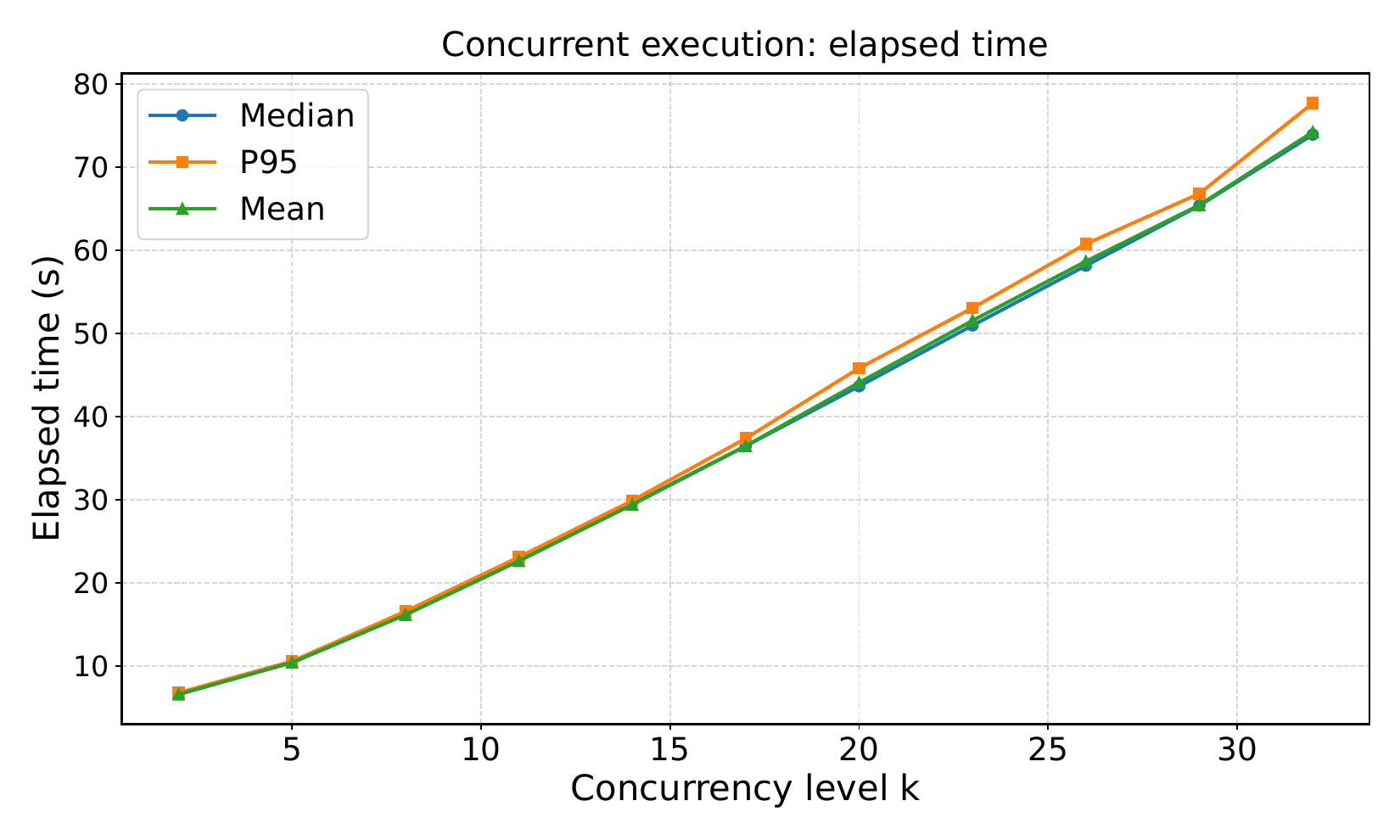}
}

\vspace{-0.8em}

\subfloat[Effective stage throughput.\label{fig:concurrent-throughput-rate}]{
    \includegraphics[width=0.96\columnwidth]{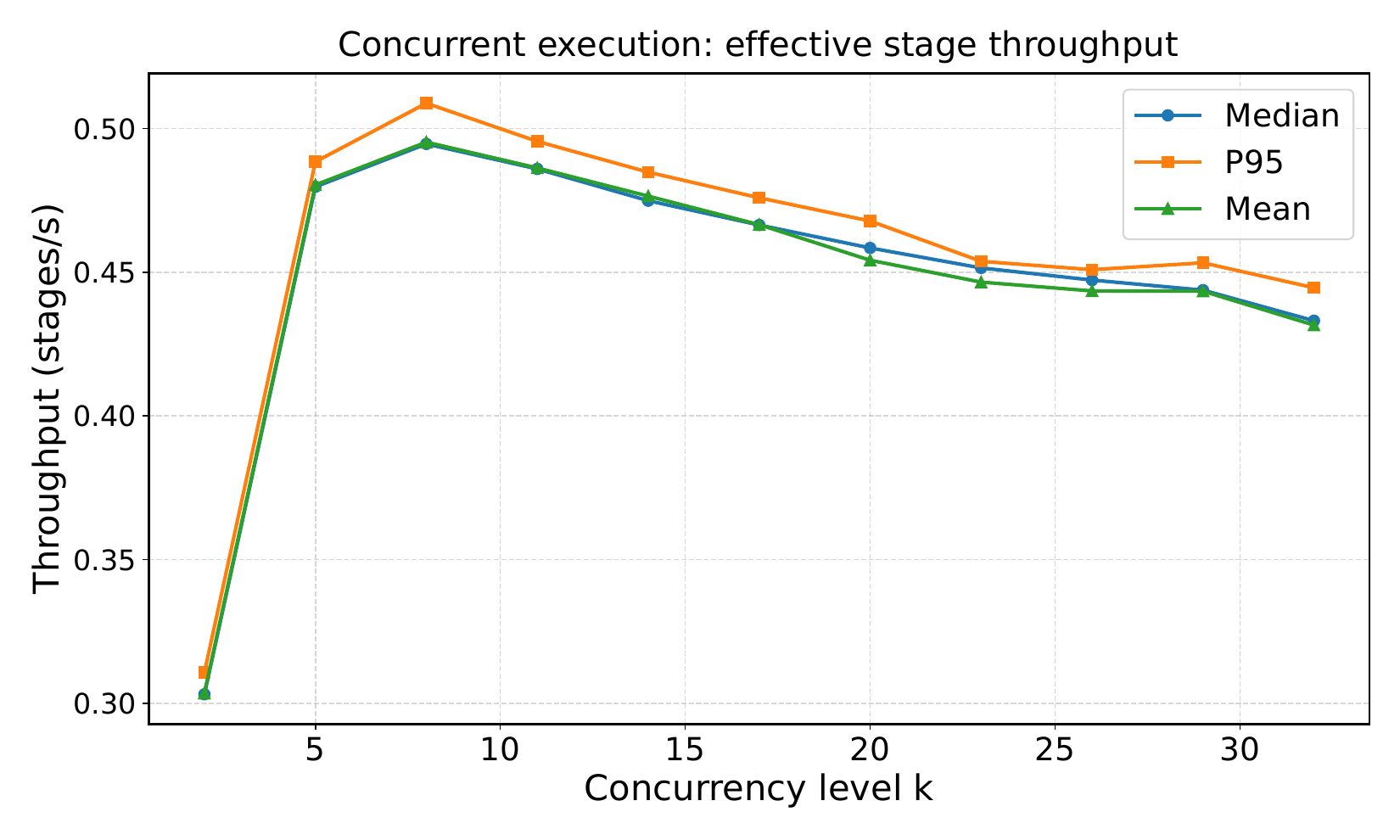}
}

\caption{Concurrent execution behavior under increasing concurrency levels. The upper panel shows total completion time, and the lower panel shows effective stage throughput.}
\label{fig:concurrent-throughput}
\end{figure}

Figure~\ref{fig:concurrent-throughput} evaluates EBCC under increasing concurrency levels. In this experiment, a \emph{stage} denotes one EBCC-managed confidential task invocation. Concretely, the request carries a \texttt{stage} field, the \texttt{serve} loop claims the request, allocates a new EID, maps the stage name to a Keystone example, executes it through the backend, and then writes the corresponding response, \texttt{meta.json}, and \texttt{run.log}. Therefore, the concurrency level $k$ means that $k$ such managed confidential stages are issued and completed in one experimental round.

The upper panel shows the total elapsed time of each round. As expected, the elapsed time increases with $k$, because more stage requests need to be claimed, dispatched, executed, and finalized. The median elapsed time grows from about 6.60~s at $k=2$ to about 73.89~s at $k=32$. All tested rounds complete successfully with a 100\% success rate, which indicates that EBCC can sustain the tested concurrency range without failed stage executions.

The lower panel reports the effective stage throughput, computed as $k / \textit{elapsed\_s}$, with the unit of stages/s. This metric is not the real execution time of an individual stage; instead, it measures how many EBCC-managed confidential stages are completed per second under a given concurrency level. The median throughput increases from about 0.30~stages/s at $k=2$ to about 0.48--0.50~stages/s at $k=5$--$8$, showing that moderate concurrency improves effective stage processing by allowing multiple managed executions to make progress within the same round. After that, the throughput gradually decreases, reaching about 0.43~stages/s at $k=32$. This suggests that higher concurrency introduces additional coordination, scheduling, and backend execution pressure, so the benefit of further increasing $k$ becomes limited.

Overall, the concurrent execution results show that EBCC's host-side coordination path remains stable under multi-stage execution. Total elapsed time naturally grows with the number of concurrent stages, but the effective throughput first improves under moderate concurrency and then degrades only gradually at higher concurrency levels. This behavior indicates that EBCC does not collapse under concurrent confidential-stage invocations; instead, its OCI-compatible management path can sustain multiple TEE-backed stages while keeping completion behavior controlled.

\subsection{Footprint and TCB Analysis}
\label{subsec:footprint-tcb}

EBCC is designed to keep OCI-facing lifecycle management and container-compatibility logic outside the enclave. In this section, we evaluate the additional footprint introduced by the Keystone-based EBCC prototype, rather than the baseline Keystone/QEMU/OCI stack or the workload's own confidential task code. The purpose of this analysis is to clarify where the prototype adds state and logic, and whether these additions enlarge the enclave-side TCB.

\begin{table}[t]
\centering
\footnotesize
\setlength{\tabcolsep}{4pt}
\renewcommand{\arraystretch}{1.05}
\caption{Measured footprint of EBCC-added host-side management components and persistent artifacts.}
\label{tab:footprint_tcb}
\begin{tabular}{llrrrr}
\toprule
\textbf{Component} & \textbf{Metric} & \textbf{Mean} & \textbf{Unit} & \textbf{Min} & \textbf{Max} \\
\midrule
Anchor     & RSS         & 1.814 & MB & 1.764 & 1.912 \\
Anchor     & VSZ         & 2.772 & MB & 2.772 & 2.772 \\
OCI Bundle & Bundle Size & 5.245 & MB & 5.245 & 5.245 \\
CID State  & State Size  & 0.016 & MB & 0.016 & 0.016 \\
\bottomrule
\end{tabular}

\vspace{0.3em}
\parbox{0.95\linewidth}{\footnotesize
\textit{Note:} RSS = resident set size; VSZ = virtual memory size; Bundle Size = on-disk OCI bundle footprint; State Size = on-disk persistent per-CID state footprint. The table reports EBCC-added management footprint in the evaluated prototype and excludes the baseline Keystone/QEMU/OCI stack and workload-intrinsic confidential task code.}
\end{table}

Table~\ref{tab:footprint_tcb} shows that the footprint added by EBCC is mainly concentrated on host-side management components and persistent artifacts. The anchor process has a small resident memory footprint, with an average RSS of 1.814~MB and a stable VSZ of 2.772~MB. The on-disk OCI bundle used by the EBCC-managed instance is about 5.245~MB, while the persistent per-CID state is only about 0.016~MB. These measurements indicate that the prototype does not require large additional memory or persistent state to maintain OCI-compatible lifecycle information.

The measured footprint also reflects the implementation structure of EBCC. The main fixed cost comes from the shared host-side control path and the OCI bundle, while the incremental per-instance state remains small. Per-stage artifacts, such as request records, response files, metadata, logs, and evidence records, grow with the number of executed stages. However, these artifacts are lightweight host-side management records: they bind a Keystone-backed stage execution to a CID-level lifecycle record and make the execution observable, but they do not duplicate the Keystone substrate or move container-management logic into the enclave.

From the TCB perspective, the key point is that EBCC does not place OCI parsing, lifecycle bookkeeping, request scheduling, stream handling, log management, or artifact persistence inside the Keystone enclave. The enclave side contains only the Keystone trusted substrate, the minimal runtime support, and the stage payload needed for confidential computation. Host-side components, including the anchor, lifecycle records, request/response artifacts, logs, and evidence files, remain outside the enclave-side TCB. They support management, coordination, recovery, and inspection, but they are not trusted to protect secret computation.

Although the measurements are taken from the Keystone prototype, they reflect EBCC's general design principle: OCI lifecycle handling and container-compatibility logic are kept on the host side, while the protected side contains only the backend-specific support needed for confidential-stage execution. EBCC therefore adds the host-side state required to make TEE-backed stages manageable through a standard container lifecycle, without materially enlarging the enclave-side TCB in the evaluated implementation.

\subsection{Security Discussion}
\label{subsec:security-discussion}

EBCC introduces host-side runtime artifacts and a stage-oriented interaction path, so it is important to clarify their security implications. EBCC does not treat host-side artifacts as trusted security objects. Files such as \texttt{state.json}, \texttt{session.json}, request and response files, \texttt{meta.json}, and \texttt{run.log} are used to preserve OCI-compatible lifecycle continuity, debugging information, and evidence visibility. They are maintained in the untrusted environment and may be observed, delayed, deleted, or corrupted by a compromised host. Therefore, EBCC does not rely on these artifacts alone to protect secrets or to authenticate confidential computation. Sensitive data and protected execution remain inside the TEE backend, and trust in a confidential stage must ultimately come from the backend's isolation and measurement or attestation mechanism.

The main additional risk introduced by EBCC is not direct disclosure of TEE memory, but host-mediated semantic interference. Because the host coordinates CID-level lifecycle state and EID-level stage execution, a compromised host may attempt to misroute a request to the wrong stage, replay an old request, bind a response to the wrong CID/EID, or drop a request before execution. These behaviors can cause denial-of-service or inconsistent host-visible state, which are within the power of the untrusted host. However, they should not allow the host to forge a valid protected-stage invocation for a different composite instance or to confuse the TEE backend about the intended session context.

To preserve this boundary, EBCC binds each accepted stage request to the current C4 context. A valid request includes the stage name, CID-related session context, epoch, sequence number, request identifier, nonce, response path, and authentication tag. The service loop validates these fields before dispatching the backend execution, allocates a fresh EID for the accepted request, and records the result under the corresponding per-stage artifact directory. This realizes the channel binding invariant in Section~\ref{subsec:sys_model}: an accepted message must be instance-bound, authenticated, fresh, and order-consistent. As a result, replayed, stale, malformed, or misrouted requests are rejected before they can be dispatched as legitimate TEE-backed stage executions.

Multi-instance execution follows the same principle. EBCC separates CID-level lifecycle state from EID-level stage records. The CID determines the managed composite instance, while the EID identifies one protected-stage invocation within that instance. A compromised host may still delay or suppress some invocations, but it cannot make a correctly implemented backend accept a request from one CID as belonging to another CID without satisfying the corresponding binding and freshness checks. Therefore, EBCC's host-side routing logic is treated as untrusted coordination rather than as part of the protected TCB.

Finally, EBCC's design does not weaken the native security boundary of the underlying TEE. For SGX, TDX, OP-TEE, or Keystone, the backend-specific isolation, measurement, and attestation mechanisms remain responsible for protecting the confidential computation. EBCC adds an OCI-compatible management layer around these mechanisms, but keeps lifecycle bookkeeping, stream handling, artifact persistence, and container compatibility logic outside the protected side. This means EBCC may increase the amount of untrusted management state, but it does not expand the trusted computation boundary with OCI control-plane logic. The remaining risks are mainly availability loss, stale or missing evidence, and incorrect host-visible orchestration state under host compromise; these are consistent with the threat model and do not constitute a break of TEE-side confidentiality or integrity.

% =========================================================
% Case Study: Feasibility on SGX / TDX / OP-TEE
% (Drop-in LaTeX section; adjust numbering as needed)
% =========================================================

\section{Case Study: Portability to Other TEEs}\label{sec:case-study}

This section studies whether EBCC's \emph{OCI-managed C4} abstraction can be adapted to representative TEEs beyond Keystone. Rather than being exhaustive, we select three TEEs that span distinct isolation granularity and invocation models: \textbf{Intel SGX} (process-level enclaves), \textbf{Intel TDX} (VM/TD-level isolation), and \textbf{ARM TrustZone/OP-TEE} (embedded secure world). Our goal is to validate that EBCC's management semantics can be preserved while replacing only the TEE-specific \emph{execution backend}, and to provide minimal but concrete evidence via small end-to-end workloads.

These three TEEs cover complementary dimensions that dominate real-world system design: (i) \textbf{isolation granularity} (process/enclave vs.\ VM/TD vs.\ secure world), (ii) \textbf{invocation and I/O path} (ECALL/OCALL vs.\ vsock/RPC into TD vs.\ TEE Client API), and (iii) \textbf{evidence and identity} (quote/report/TA identity). This selection allows us to demonstrate EBCC's portability across qualitatively different TEEs without inflating scope.

\subsection{Model-Grounded Mapping: ``TEE Ops'' to EBCC State Transitions}
\label{subsec:cs-mapping}

EBCC abstracts confidential execution as two coupled state machines: an \emph{instance-level} lifecycle
(Created $\rightarrow$ Running $\rightarrow$ Stopped $\rightarrow$ Deleted) and a \emph{stage-level} execution pipeline
(ReqPending $\rightarrow$ Claimed $\rightarrow$ Prepared $\rightarrow$ Executing $\rightarrow$ Completed/Failed).
Porting EBCC to a new TEE amounts to instantiating the \emph{Prepared/Executing} transitions with TEE-specific primitives,
while keeping the CID lifecycle, request claiming (exactly-once), and evidence persistence unchanged.
For completeness, Appendix~\ref{app:mapping} provides the full transition-to-primitive mapping in Table~\ref{tab:full-mapping}.

% (Optional) Keep a compact summary table here in main text,
% and move the full, wide mapping table to Appendix as a rotated landscape table.

\subsection{Backend Adapter Interface and Evidence Artifacts}
\label{subsec:cs-adapter}

To avoid changing EBCC's core management logic for each TEE backend, we isolate backend-specific details behind a small adapter interface. The \texttt{serve} loop remains responsible for request claiming, EID allocation, lifecycle-state updates, evidence-artifact persistence, and response generation. It delegates only the backend-dependent protected execution step to the adapter:

\begin{verbatim}
prepare(CID, EID, stage) -> handle
execute(handle, request) ->
       (rc, stdout, evidence)
destroy(handle)
\end{verbatim}

The \texttt{prepare} call creates or binds the backend-specific protected execution context for a given CID, EID, and stage, and returns an opaque \texttt{handle}. This handle may correspond to different concrete objects on different TEEs, such as an enclave instance, a confidential-VM execution context, or a TA session. The \texttt{execute} call submits the stage request to that protected context and returns the execution result, including the return code, standard output, and backend-specific evidence. The \texttt{destroy} call releases the backend context after the stage finishes, without exposing backend-specific cleanup details to EBCC's core management logic.

\noindent\textbf{Evidence artifacts.}
For all backends, EBCC writes per-stage evidence artifacts under
\texttt{StateDir/<CID>/enclaves/<EID>/}, such as \texttt{meta.json} and \texttt{run.log},
and records summary fields in \texttt{state.json}, such as \texttt{last\_stage}, \texttt{last\_rc}, and \texttt{last\_eid}.
Backend-specific evidence, such as a quote, report, measurement, or TA identity, is stored as additional fields in
\texttt{meta.json}, including fields such as \texttt{tee\_type}, \texttt{evidence\_type}, and
\texttt{measurement\_hash}. In this way, EBCC keeps the lifecycle and artifact layout uniform while allowing each TEE backend to provide its own native evidence format.

\subsection{Backend Feasibility: SGX, TDX, and OP-TEE}
\label{subsec:cs-backends}

\paragraph{SGX backend (process-level enclaves).}
SGX naturally fits EBCC's stage abstraction: each stage corresponds to an enclave invocation session.
The backend maps \emph{Prepared} to enclave creation/initialization (or enclave pooling), and \emph{Executing} to ECALL.
Upon completion, EBCC persists \texttt{run.log} and updates \texttt{meta.json} with \texttt{tee\_type=sgx}, return code,
and optional measurement/quote status.

\paragraph{TDX backend (VM/TD-level isolation).}
TDX changes isolation granularity from an enclave to a protected VM (TD).
The backend maps \emph{Prepared} to ensuring a TD-resident runner endpoint is reachable and binding a communication session
to the EID, and maps \emph{Executing} to an RPC/vsock request into the TD and collecting the response.
EBCC records \texttt{tee\_type=tdx} and channel/session identifiers (and optional TD report status) in \texttt{meta.json}.

\paragraph{OP-TEE backend (embedded secure world).}
OP-TEE runs trusted applications (TAs) in the secure world invoked from the normal world via the TEE Client API.
The backend maps \emph{Prepared} to \texttt{TEEC\_OpenSession} (binding TA UUID and shared memory to EID),
and maps \emph{Executing} to \texttt{TEEC\_InvokeCommand} (with a stage-defined command id).
EBCC records \texttt{tee\_type=optee}, TA UUID/cmd id, and return origin in \texttt{meta.json}.
For repeatability, we use QEMU-based OP-TEE emulation.

\subsection{Cross-TEE Workload Evaluation}
\label{subsec:cs-miniexp}

We further conduct cross-TEE workload evaluations on three different TEE backends, namely SGX, TDX, and OP-TEE. The purpose of this experiment is not to compare the absolute performance of different TEEs, because they represent different hardware abstractions, execution granularities, and software stacks. Instead, the goal is to validate whether the same EBCC lifecycle abstraction and stage pipeline can be instantiated across enclave-style, VM-style, and embedded-style TEEs.

Each backend executes two representative workloads. The light task is a \texttt{hello} workload. On SGX, it enters an enclave and returns a short string; on TDX, it is dispatched to a command inside the confidential VM; on OP-TEE, it invokes a TA command and returns a message from the trusted world. The heavy task is an AES-128-GCM workload over 16~MiB of logical input. The concrete implementation is backend-specific: SGX performs the encryption inside the enclave, TDX dispatches the \texttt{aesgcm16m} stage inside the confidential VM, and OP-TEE invokes a crypto TA that generates the input stream inside the TA and returns the authentication tag. For each backend, we compare native execution with EBCC-managed execution and report the end-to-end latency of first-stage completion, including median, P95, and mean latency over 50 repetitions.

\begin{figure}[!t]
    \centering
    \includegraphics[width=\columnwidth]{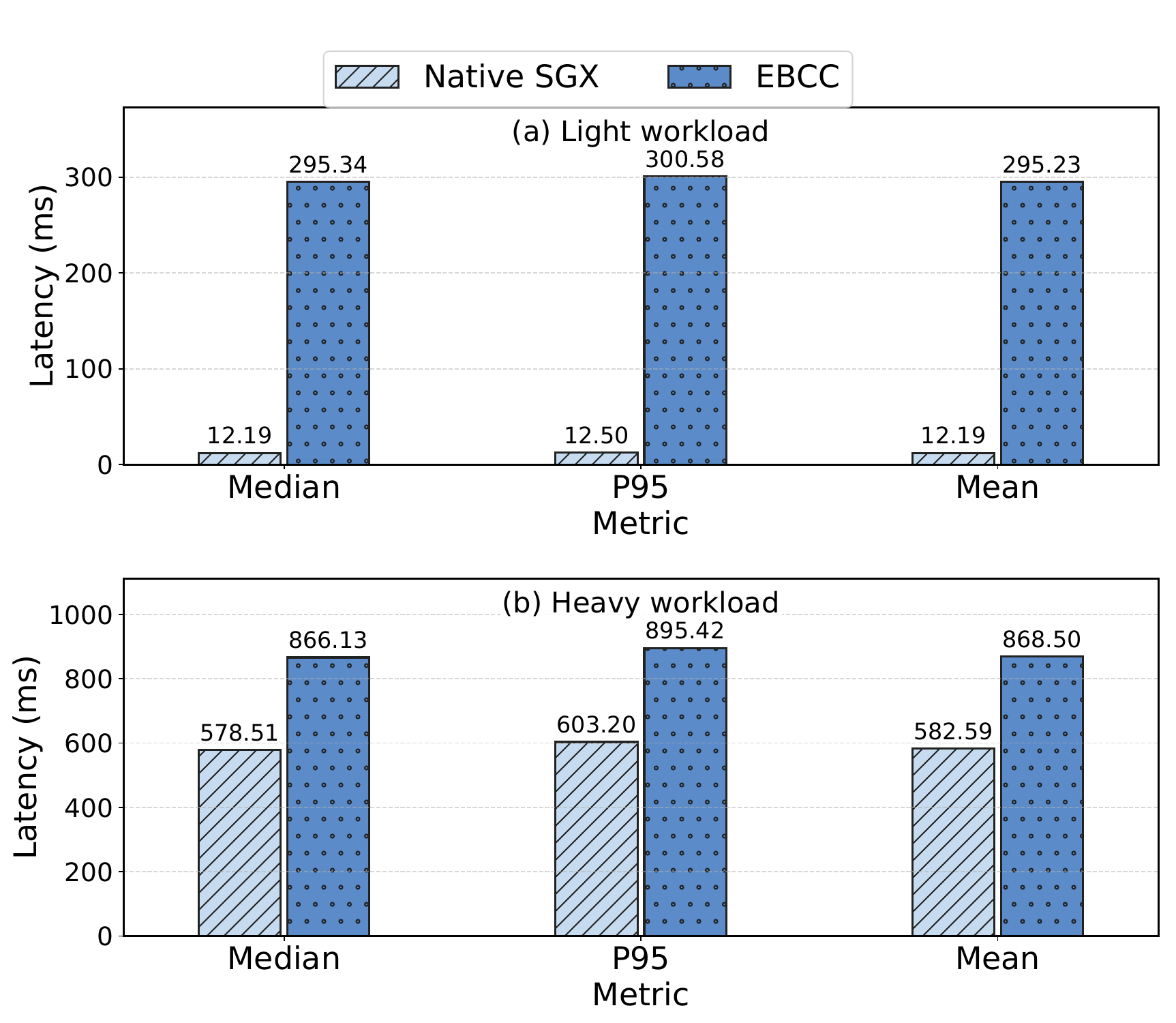}
    \caption{Workload evaluation on SGX. The upper panel shows the light task, and the lower panel shows the heavy task.}
    \label{fig:cs-sgx}
\end{figure}

\begin{figure}[!t]
    \centering
    \includegraphics[width=\columnwidth]{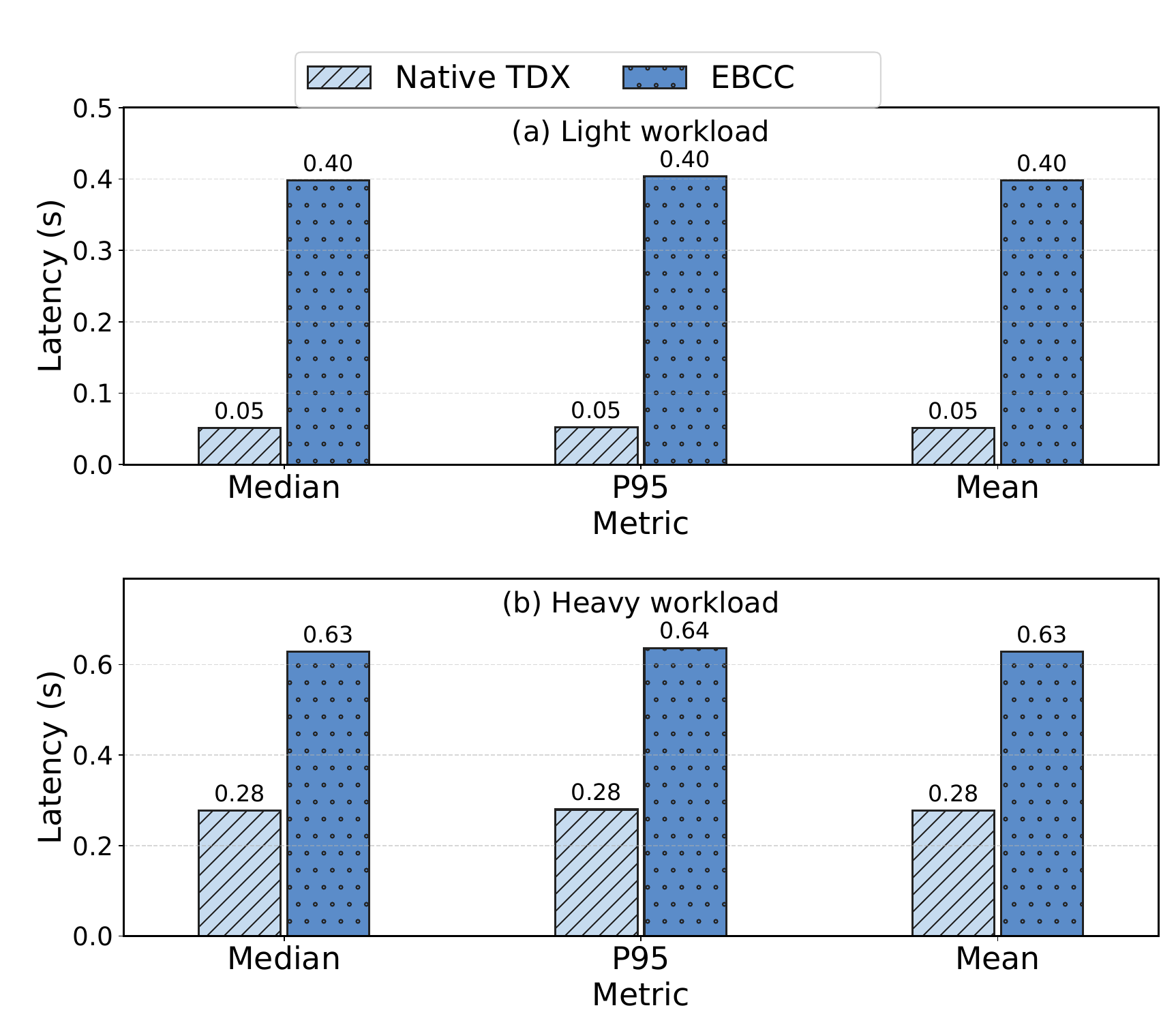}
    \caption{Workload evaluation on TDX. The upper panel shows the light task, and the lower panel shows the heavy task.}
    \label{fig:cs-tdx}
\end{figure}

\begin{figure}[!t]
    \centering
    \includegraphics[width=\columnwidth]{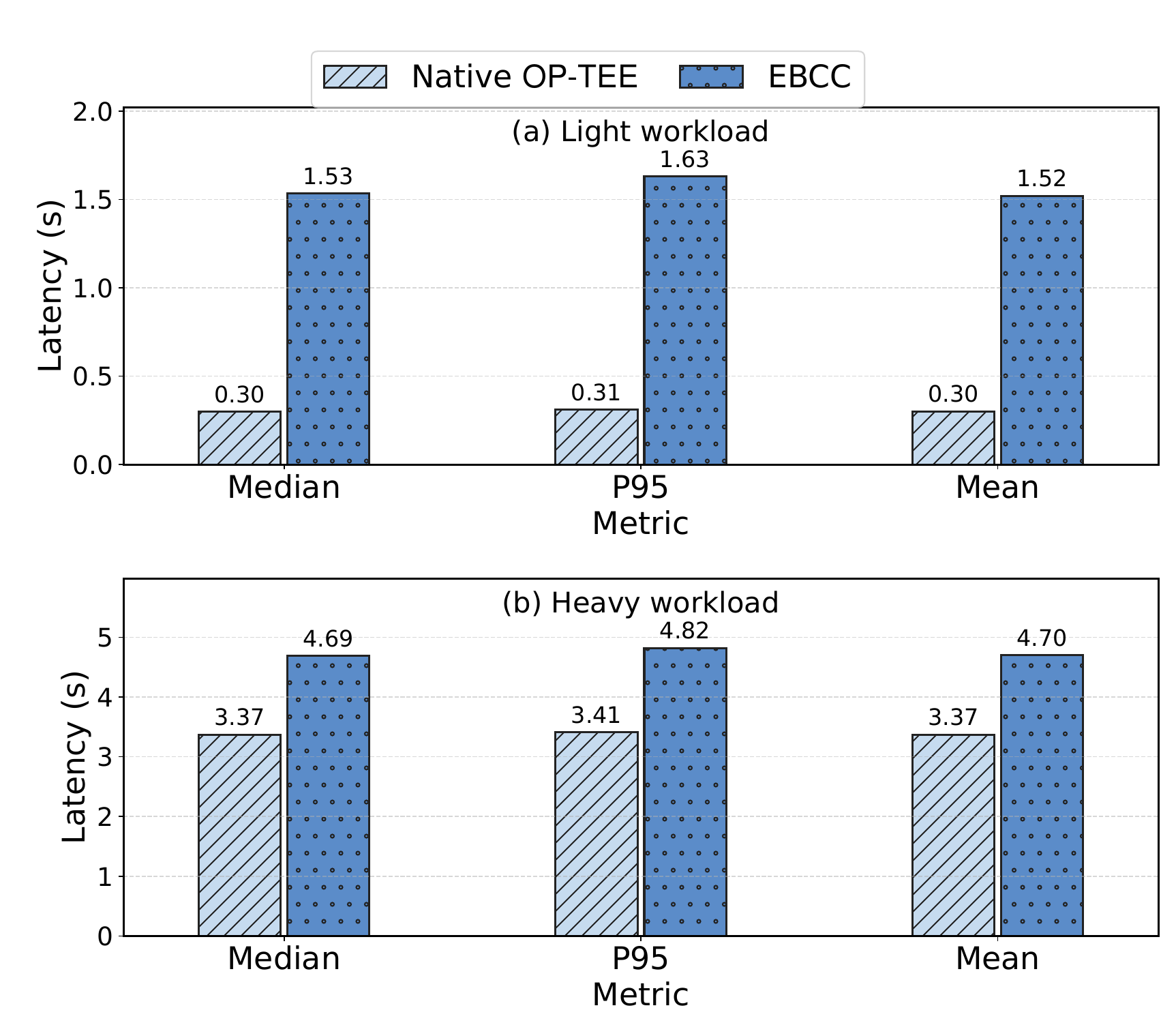}
    \caption{Workload evaluation on OP-TEE. The upper panel shows the light task, and the lower panel shows the heavy task.}
    \label{fig:cs-optee}
\end{figure}

Figure~\ref{fig:cs-sgx} shows the SGX case. For the light workload, native SGX completes in about 12.19~ms at the median, while EBCC-managed execution takes about 295.34~ms. For the heavy AES-128-GCM workload, native SGX increases to about 578.51~ms, and EBCC-managed execution reaches about 866.13~ms. The absolute EBCC overhead is therefore similar in the two workloads, but its relative impact is much larger for the light task. This is consistent with the SGX adapter path: EBCC adds request claiming, EID allocation, stage dispatching, artifact generation, and response synchronization around native SGX execution. When the protected computation is short, this managed path dominates the total latency; when the AES workload becomes longer, the same fixed management cost is partially amortized by the protected SGX computation.

Figure~\ref{fig:cs-tdx} reports the TDX case. Native TDX completes the light workload in about 0.05~s and the heavy workload in about 0.28~s at the median. Under EBCC-managed execution, the corresponding median latencies are about 0.40~s and 0.63~s. Unlike SGX, the TDX backend follows a VM-style execution model: EBCC sends \texttt{prepare}, \texttt{execute}, and \texttt{destroy}-style requests to a runner inside the confidential VM, and the runner binds the stage to a session, channel identifier, measurement hash, and report digest. The similar EBCC--native gap in both light and heavy workloads indicates that the dominant additional cost is the managed invocation and evidence-handling path across the VM boundary, rather than the AES computation itself.

Figure~\ref{fig:cs-optee} shows the OP-TEE case. For the light workload, native OP-TEE completes in about 0.30~s, while EBCC-managed execution takes about 1.54~s at the median. For the heavy AES-128-GCM workload, native OP-TEE takes about 3.37~s, and EBCC-managed execution takes about 4.69~s. The OP-TEE backend uses the GlobalPlatform-style CA/TA interaction: the light task opens a TA session and invokes a hello command, while the heavy task invokes a crypto TA that performs AES-GCM inside the trusted world and returns only the final tag. The EBCC overhead remains visible, but the gap is again less dominant for the heavy workload because the trusted-world computation occupies a larger fraction of the total execution time.

Across the three case studies, the absolute latency differs substantially because SGX, TDX, and OP-TEE expose different execution granularities and backend paths. SGX uses process-level enclave creation and ECALL execution; TDX uses VM-granularity protected execution and an RPC-style runner inside the TD; OP-TEE uses CA/TA sessions and trusted-world command invocation. EBCC does not hide these backend costs, nor does it aim to make the three TEEs perform similarly. The important observation is that all three backends can be driven through the same EBCC-managed structure: a host-side anchor emits a stage request, the EBCC runtime claims and validates it, allocates an EID, dispatches the backend-specific protected execution, records metadata and logs, and returns a response through the same lifecycle-visible path.

These results support two conclusions. First, EBCC is practically adaptable to heterogeneous TEEs, including enclave-style SGX, VM-style TDX, and embedded-style OP-TEE. The backend-specific execution details are confined to the adapter layer, while the upper-level lifecycle and artifact structure remain unchanged. Second, EBCC introduces additional latency compared with native execution, but this overhead corresponds to useful management and security semantics: request binding, per-stage identification, response synchronization, and evidence-artifact persistence. Thus, the case studies confirm that EBCC can preserve a unified execution interface and availability across heterogeneous TEEs while maintaining the intended security boundary of each backend.

\subsection{Discussion}
\label{subsec:cs-discussion}

The case studies clarify the portability boundary of EBCC. EBCC does not abstract away the performance characteristics or security assumptions of individual TEEs; SGX, TDX, and OP-TEE still expose different invocation costs, evidence formats, and resource constraints. What EBCC unifies is the management layer above them: CID-level lifecycle control, EID-level stage execution, request/response synchronization, and durable artifact generation. This distinction is important because portability here means that heterogeneous TEEs can be connected through the same stage interface without changing the OCI-facing lifecycle logic, not that they provide identical performance or identical attestation semantics. The measured overhead is therefore the cost of preserving this managed execution structure, while the successful completion of both workloads across all three backends shows that the abstraction is usable beyond the Keystone prototype.

\section{Related Work}
\label{sec:related-work}

\paragraph{Confidential containers and cloud-native confidential computing}
Confidential containers have become an important direction for bringing TEE protection into cloud-native environments. CoCo and related studies analyze or extend VM-backed confidential-container stacks from the perspectives of security, deployment, attestation, and performance, showing both the practicality of confidential-container execution and the complexity of the surrounding control plane \cite{coco_ccs24_shiftedmoats,segarra2024serverlesscoco}. Other work further studies privacy-preserving attestation, confidential-Kubernetes deployment models, and attestation-/KBS-based secret-release workflows \cite{privacy_preserving_container_attestation_2026,confidential_k8s_models_2025,cncf_coco_self_assessment,coco_attestation_agent,coco_trustee_kbs,coco_attestation_service_repo,coco_attestation_docs_2025}. These systems are important because they demonstrate how confidential computing can be operationalized in cloud-native stacks. EBCC addresses a complementary boundary: rather than building a full confidential-container platform, it asks how a composite REE--TEE workload can preserve OCI-managed lifecycle semantics while keeping TEE-specific execution behind a stage-level backend adapter.

\paragraph{TEE-backed container architectures and trust-boundary engineering}
Several systems explore how container abstractions should be combined with trusted hardware. Trusted Container Extensions study container-level confidential-computing support, while RContainer extends ARM CCA primitives to build a secure container architecture \cite{tcx_2022,rcontainer_ndss25}. Our recent Arca work explores a TEE-in-Container design point that keeps orchestration logic outside the TEE and reduces enclave-side footprint \cite{arca2026}. Related studies also point out that TEE-container systems introduce new trust-boundary engineering problems: container control paths, host-side orchestration, and developer-facing APIs may create vulnerabilities or misuse risks even when the underlying TEE is sound \cite{tcon_trust_boundary_vuln_2025,misuse_tee_2025,sok_tee_usage_ccsw23}. EBCC follows the same principle of avoiding unnecessary protected-side expansion, but focuses on a narrower runtime question: keeping OCI lifecycle handling outside the TEE while binding each confidential stage to durable state and evidence artifacts.

\paragraph{Executing applications inside TEEs}
A second line of work focuses on how to run applications inside specific TEEs. SCONE, Occlum, and Graphene-SGX provide library-OS or compatibility-layer support so that container-style or largely unmodified applications can run over SGX enclave execution substrates \cite{scone_osdi16,occlum_asplos20,graphene_sgx_atc17}. Later systems and studies extend this direction to newer TEE settings and practical workloads, including VM-style TEEs such as TDX \cite{gramine_tdx_2024,miwa2023gramine_hpc}. Other efforts explore reusable enclaves, middleware abstractions, and WebAssembly-oriented trusted runtimes \cite{acctee_middleware19,tian_codaspy19_sgx_containers,twine_arxiv21,chancel_2021,reusable_enclaves_usenix23}. These works mainly improve the in-TEE execution substrate or compatibility layer. EBCC is complementary: it does not try to define a new in-TEE application runtime, but instead defines how TEE-backed stages can be invoked, observed, and coordinated through an OCI-compatible lifecycle.

\paragraph{VM-based TEEs and heterogeneous confidential-computing backends}
VM-based TEEs such as TDX, SEV-SNP, and ARM CCA shift the granularity of trusted execution from enclave processes to protected VMs or realms, changing invocation paths, measurement semantics, and host--guest trust boundaries. Prior studies clarify the TDX platform model and compare TDX and SEV-SNP in terms of performance and security properties \cite{cheng2024tdxdemystified,intel_tdx_docs,empirical_sevsnp_tdx_2024,coppolino2025tee_eval}. ARM CCA and related systems further extend the confidential-computing design space to realm-based execution and accelerator-aware confidential computing \cite{armcca_login2023,cage_ndss2024}. Verified CCAaaS work studies how confidential-computing services can be specified and verified at a higher assurance level \cite{verified_ccaas_usenix23}. These efforts show that confidential execution is increasingly heterogeneous. EBCC uses this observation as a design constraint: the upper-level lifecycle and artifact structure should remain stable, while backend-specific execution details are isolated behind an adapter.

\paragraph{TEE backend threat surfaces}
Backend TEEs cannot be treated as transparent trusted black boxes. A substantial body of work shows that privileged control paths, VM interfaces, interrupt behavior, and microarchitectural side channels can weaken or complicate confidential-computing guarantees \cite{heckler_usenix24,wesee_sp24,00seven_usenix24,cachewarp_usenix24,ciphertext_side_channels_usenix21,rmpocalypse_ccs25,heracles_ccs25,counterseveillance_ndss25,tdxdown_ccs24,tdxploit_usenix25,sevstep_tches2024,tdx_sidechannels_iacr2025}. Complementary enclave-oriented analyses reveal similar sensitivities in interrupt handling and orchestration-facing control paths \cite{pandora_sp24,aex_notify_usenix23,aex_nstep_arxiv25,sigy_asiaccs2025}. For EBCC, these studies motivate an explicit separation between lifecycle management and backend-specific protected execution. EBCC does not assume that a TEE backend is a simple drop-in execution target; instead, it makes stage transitions, request binding, backend evidence, and host-side coordination explicit.

\paragraph{Attestation, evidence, and verifier-side trust frameworks}
Attestation and evidence frameworks define how measured execution is represented and consumed by verifiers. The IETF RATS architecture and related standards define attestation roles and evidence formats, while EAT and PSA-token standards provide interoperable representations for measured claims \cite{ietf_rats_arch_draft,rfc9334_rats_architecture,rfc9711_eat,rfc9782_eat_media_types,rfc9783_psa_token}. Verifier frameworks and protocol studies further address standards-based verification, attestation communication, constrained disclosure, and trust-decision workflows \cite{veraison_2024,provably_secure_ra_protocols_2024,prove_sciencedirect_2023,universal_ra_2023,racd_acsac23}. Systems work also studies operational issues such as continuous integrity attestation, update integration, and self-verifying attestation evidence \cite{keylime_dsn25_case_study,rasues_2026,integrity_measurement_cc_2026,trustmee_arxiv26}. These efforts define how evidence can be generated, encoded, verified, and used, but they do not by themselves specify how a composite REE--TEE workload should be driven through an OCI lifecycle. EBCC is complementary: it records backend-specific evidence as durable per-stage artifacts and binds them to lifecycle-visible execution state.

\paragraph{Summary}
Existing work mainly addresses four layers: building confidential-container stacks and their attestation/deployment ecosystem, providing execution substrates for applications inside specific TEEs, analyzing backend threat surfaces and trust-boundary pitfalls, and defining evidence formats or verifier-side trust frameworks. EBCC targets a different layer. It preserves OCI-managed lifecycle semantics for a composite execution instance, keeps TEE-specific execution behind a backend adapter, and exposes stage-level results and evidence as durable artifacts. This makes TEE-backed execution manageable through an OCI-style lifecycle without requiring the OCI control path itself to be moved into the TEE.

\section{Conclusion}\label{sec:concl}

This paper presented \SYSNAME, an OCI-compatible runtime architecture for managing C4 confidential-computing composites. The central idea is to treat a confidential-computing workload as a composite execution instance: the REE-side anchor remains responsible for the surrounding business logic and external interaction, while TEE-side confidential stages are invoked on demand through a stage-level backend adapter. By preserving the OCI runtime boundary, \SYSNAME\ allows container managers to control such composite workloads through standard lifecycle operations, while the runtime maintains the persistent CID-level state, EID allocation, request/response paths, metadata, logs, and evidence artifacts needed to make TEE-backed execution observable and manageable.

We implemented \SYSNAME\ on Keystone and evaluated its lifecycle correctness, runtime overhead, footprint, and concurrent execution behavior. The results show that EBCC's additional cold-start cost is mainly concentrated in the \textit{create} phase, where the runtime prepares per-instance state and execution context, while the \textit{start} phase remains small. End-to-end measurements show that EBCC is slower than native Keystone, but the gap corresponds to the additional managed execution path required for request validation, replay protection, EID assignment, backend dispatch, artifact persistence, and response synchronization. The footprint analysis further shows that the added state is concentrated on the host side and does not materially enlarge the protected-side TCB.

We also studied the portability of the design through SGX, TDX, and OP-TEE case studies. These backends expose different execution models---process-level enclaves, VM-style confidential execution, and embedded trusted-world execution---yet all can be connected to EBCC through the same lifecycle and stage abstraction. The case studies confirm that backend-specific details can be confined to adapter logic, while the upper-level management path remains stable. Overall, EBCC trades additional runtime latency for a unified, OCI-compatible way to manage confidential stages, preserve evidence artifacts, and keep TEE-specific execution behind a narrow backend interface. Future work will further integrate production-grade attestation and secret provisioning, strengthen artifact verification policies, and explore richer multi-stage confidential workloads across heterogeneous TEE deployments.

\section*{Acknowledgments}
The authors would like to thank the editor-in-chief, associate editor, and reviewers for their valuable comments and suggestions. This research was supported by the National Natural Science Foundation of China (62232013, U24A20243, 62572377, 62302363), the Innovation Capability Support Program of Shaanxi (No. 2023-CX-TD-02), the Xidian University Specially Funded Project for Interdisciplinary Exploration (No. TZJHF202502) and the Fundamental Research Funds for the Central Universities (No. ZDRC2202).


\bibliographystyle{IEEEtran}
\bibliography{refs}

\begin{IEEEbiography}[{\includegraphics[width=1in,height=1.25in,clip,keepaspectratio]{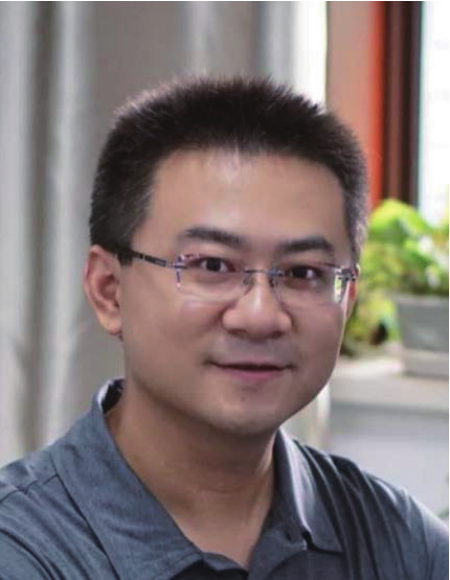}}]{Di Lu}
(Member, IEEE) received the B.S., M.S., and Ph.D. degrees in computer science and technology from Xidian University, China, in 2006, 2009, and 2014. Now he is a full  Professor in the School of Computer Science and Technology at Xidian University. His research interests include trusted computing, confidential computing, system and network security.
\end{IEEEbiography}

\begin{IEEEbiography}[{\includegraphics[width=1in,height=1.25in,clip,keepaspectratio]{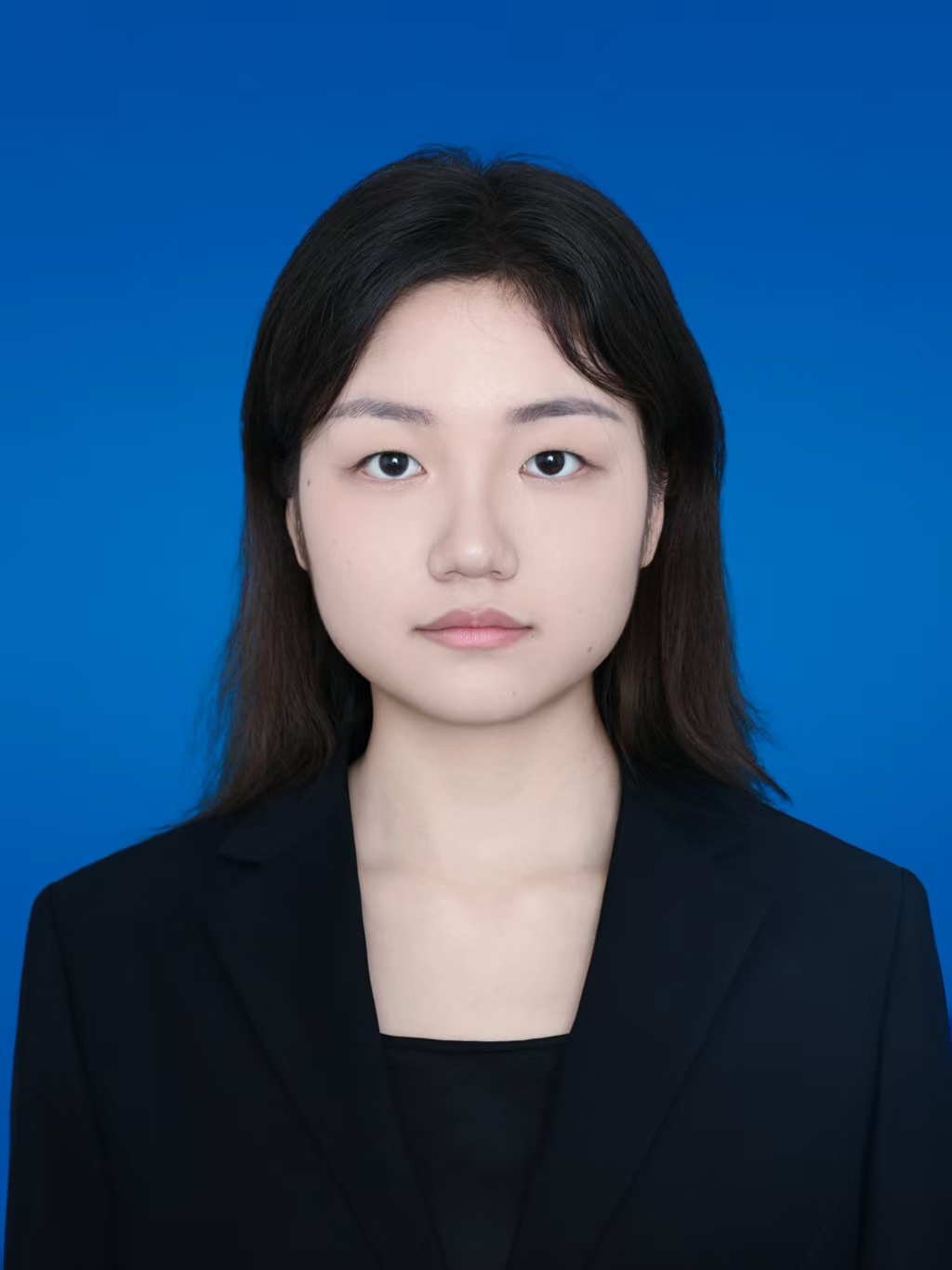}}]{Qingwen~Zhang}
received the BS degree from Xidian University, China, in 2024. She is currently pursuing an MS degree in the School of Computer Science and Technology at Xidian University, China. Her research interests include trusted computing and embedded system security.
\end{IEEEbiography}

\begin{IEEEbiography}[{\includegraphics[width=1in,height=1.25in,clip,keepaspectratio]{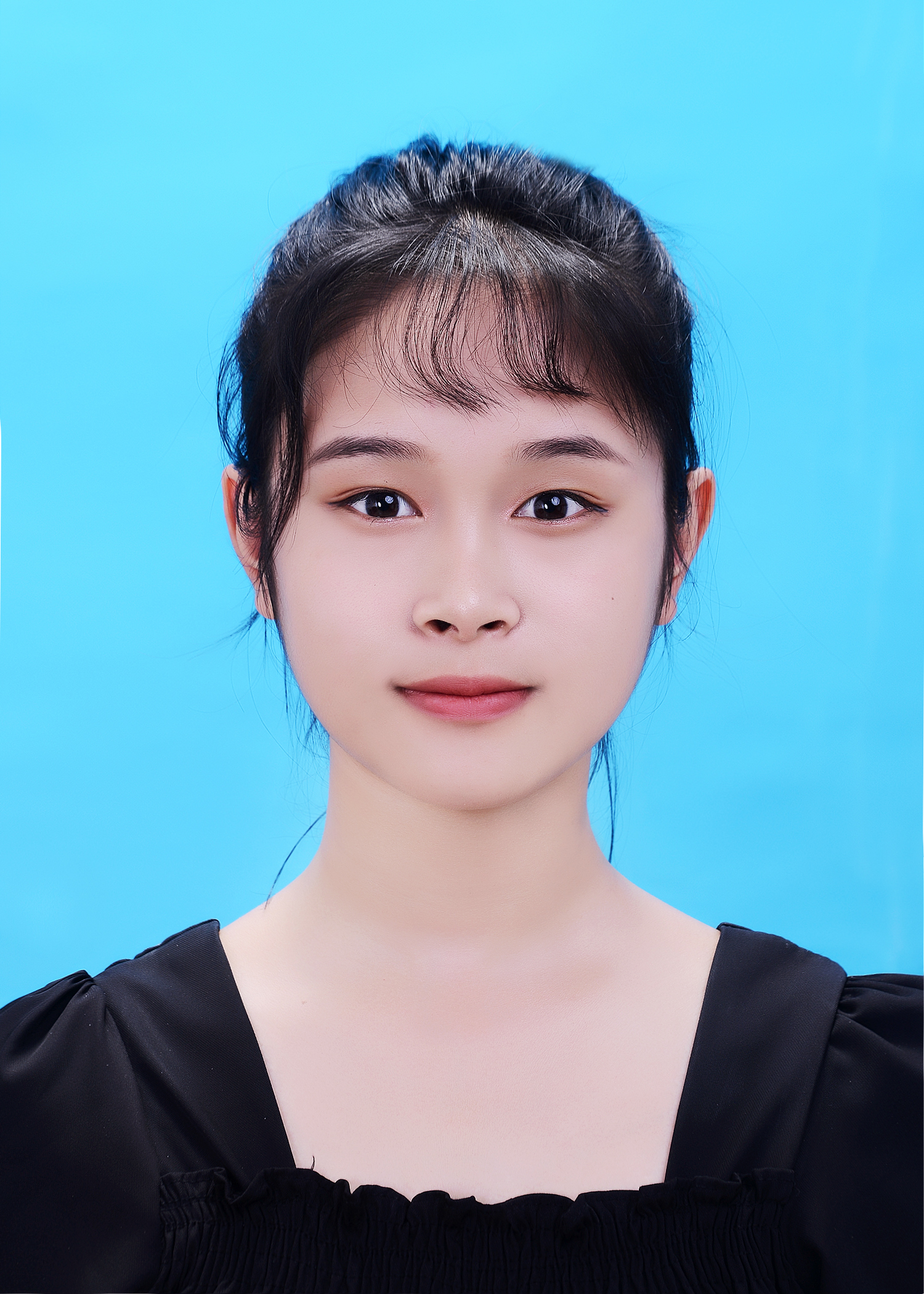}}]{Yujia~Liu} received the BS degree from Xi’an Shiyou University, China, in 2025. She is currently pursuing an MS degree in the School of Computer Science and Technology at Xidian University, China. Her research interests include trusted computing and embedded system security. 
\end{IEEEbiography}

\begin{IEEEbiography}[{\includegraphics[width=1in,height=1.25in,clip,keepaspectratio]{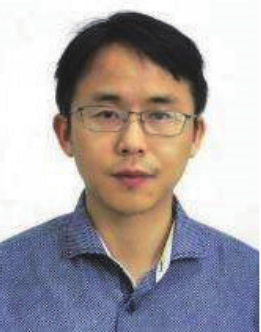}}]{Xuewen Dong}
(Member, IEEE) received the B.E., M.S., and Ph.D. degrees in computer science and technology from Xidian University, Xi’an, China, in 2003, 2006, and 2011, respectively. From 2016 to 2017, he was a Visiting Scholar with Oklahoma State University, Stillwater, OK, USA. Currently, he is a Professor with the School of Computer Science and Technology, Xidian University. His research interests include blockchain and the security of smart systems.
\end{IEEEbiography}

\begin{IEEEbiography}[{\includegraphics[width=1in,height=1.25in,clip,keepaspectratio]{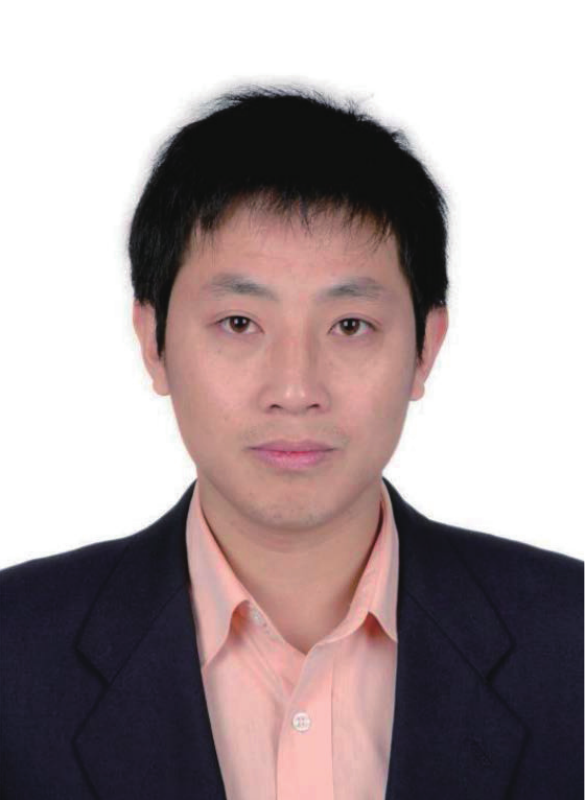}}]{YuLong Shen}
(Member, IEEE) received the BS and MS degrees in computer science and a PhD degree in cryptography from Xidian University, Xi’an, China, in 2002, 2005, and 2008, respectively. He is currently a professor with the School of Computer Science and Technology, Xidian University. His research interests
include wireless network security and cloud computing security.
\end{IEEEbiography}

\begin{IEEEbiography}[{\includegraphics[width=1in,height=1.25in,clip,keepaspectratio]{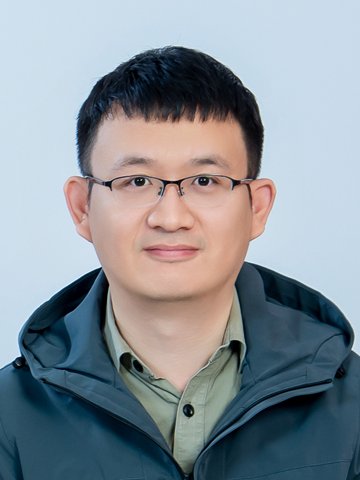}}]{Zhiquan Liu}
received the B.S. degree from the School of Science, Xidian University, Xi'an, China, in 2012, and the Ph.D. degree from the School of Computer Science and Technology, Xidian University, Xi'an, China, in 2017.

He is currently a full professor, doctoral supervisor, and deputy dean with the College of Cyber Security, Jinan University, Guangzhou, China. His current research focuses on security, trust, privacy, and intelligence in vehicular networks and UAV networks. He currently serves as the area editor or associate editor of multiple SCI-index journals, such as IEEE TIFS, IEEE TDSC, IEEE TII, IEEE TVT, IEEE IOTJ, IEEE Network, Information Fusion, etc. His homepage is https://www.zqliu.com.
\end{IEEEbiography}

\begin{IEEEbiography}[{\includegraphics[width=1in,height=1.25in,clip,keepaspectratio]{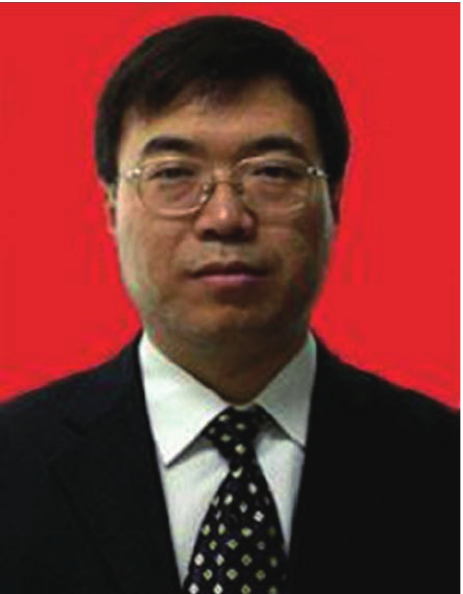}}]{Jianfeng Ma}
(Member, IEEE) received the BS degree in mathematics from Shaanxi Normal University, China, in 1985, and the MS and PhD degrees in computer software and communications engineering from Xidian University, China, in 1988 and 1995, respectively. Now, he is a professor with the School of Cyber Engineering, Xidian University, China. His current research interests include distributed systems, computer networks, and information and network security.
\end{IEEEbiography}

\clearpage
\appendices
\section{Additional System-Model Details}
\label{app:system-model-details}

This appendix provides the detailed formal components omitted from Section~\ref{subsec:sys_model}, including the minimal state-record fields, auxiliary observability semantics, detailed OCI entrypoint rules, and composite termination semantics.

\subsection{Minimal State-Record Fields}
\label{app:minimal-record}

The CID-level state record \(SR(CID)=\langle CID,S,ver\rangle\) maintains three minimal invariants:
\begin{itemize}
  \item[\(I1\)] \emph{Existence decidability:} \(\textsf{Init}\) is decidable by the absence of \(SR(CID)\).
  \item[\(I2\)] \emph{State recoverability:} the primary state \(S\) is recoverable across multi-call OCI invocations.
  \item[\(I3\)] \emph{Update orderability:} state updates are orderable via a monotone \(ver\), preventing stale writes from overwriting newer records.
\end{itemize}
The field set \(F=\{CID,S,ver\}\) is minimal for maintaining these invariants. This can be captured by a mapping
\(M:F\rightarrow\mathcal{P}(I_{\min})\), where \(I_{\min}=\{I1,I2,I3\}\):
\[
\begin{aligned}
M(CID)=\{I1\},\quad M(S)=\{I2\},\quad M(ver)=\{I3\},\\
\bigcup_{f\in F}M(f)=I_{\min}.
\end{aligned}
\]
Removing any field \(f\in F\) breaks at least one invariant: removing \(CID\) breaks instance identity and existence decidability, removing \(S\) breaks lifecycle recovery, and removing \(ver\) breaks ordered updates under multi-call or concurrent invocations.

\subsection{Auxiliary Observability and Readiness}
\label{app:observability}

The primary lifecycle state intentionally remains small. EBCC therefore attaches auxiliary observability fields that do not participate in primary state transitions:
\[
\begin{aligned}
trust\_flag &\in \{\textsf{trusted},\ \textsf{untrusted},\ \textsf{unknown}\},\\
health\_flag &\in \{\textsf{healthy},\ \textsf{degraded},\ \textsf{unknown}\},\\
tee\_phase &\in \{\textsf{idle},\ \textsf{active},\ \textsf{error}\}.
\end{aligned}
\]
These fields refine the interpretation of OCI-visible states without overloading them. For example, a C4 instance may be OCI-visible as \textsf{running} while its trust state is still \textsf{unknown}, or while no TEE-backed stage is active.

The evidence inputs for these observability fields can be modeled as:
\[
\begin{aligned}
\mathcal{E}_{trust} &:= \langle e_{\textsf{att}},\ e_{\textsf{meas}},\ e_{\textsf{bind}}\rangle,\\
\mathcal{E}_{health} &:= \langle e_{\textsf{dep}},\ e_{\textsf{res}},\ e_{\textsf{perf}}\rangle,\\
\mathcal{E}_{tee} &:= \langle e_{\textsf{call}},\ e_{\textsf{timeout}},\ e_{\textsf{exit}}\rangle.
\end{aligned}
\]
Here \(e_{\textsf{att}}\), \(e_{\textsf{meas}}\), and \(e_{\textsf{bind}}\) denote attestation, measurement, and binding evidence; \(e_{\textsf{dep}}\), \(e_{\textsf{res}}\), and \(e_{\textsf{perf}}\) summarize dependency, resource, and performance observations; and \(e_{\textsf{call}}\), \(e_{\textsf{timeout}}\), and \(e_{\textsf{exit}}\) describe TEE-stage activity and exit behavior.

Given a deployment-specific trust policy
\[
P:\mathcal{E}_{trust}\rightarrow\{0,1\},
\]
the auxiliary fields can be derived as:
\[
\begin{aligned}
trust\_flag &:= \mathsf{TrustEval}(CID;\mathcal{E}_{trust},P),\\
health\_flag &:= \mathsf{HealthEval}(CID;\mathcal{E}_{health}),\\
tee\_phase &:= \mathsf{TEEPhaseEval}(CID;\mathcal{E}_{tee}).
\end{aligned}
\]

Readiness is modeled as a derived predicate rather than as a primary lifecycle state:
\[
\mathsf{Ready}(CID)=1\Rightarrow S=\textsf{Running}.
\]
We further distinguish REE-side and TEE-side preparedness:
\[
\begin{split}
\mathsf{Prepared}_R(CID) := \mathbf{1}\!\left[
\mathsf{AnchorAlive}(CID)\wedge\mathsf{DepsOK}(CID)\wedge
\mathsf{InitDone}(CID)
\right],
\end{split}
\]
\[
\mathsf{Prepared}_T(CID) := \mathbf{1}\!\left[
\exists\,EID\in\mathcal{E}(CID):\mathsf{EReady}(EID)
\right].
\]
Then readiness is selected by policy:
\[
\begin{aligned}
\mathsf{Ready}(CID)=1 \Longleftrightarrow\ 
&\mathsf{Prepared}_R(CID)=1\\
&\wedge\big(\neg \mathsf{RequireConf}(CID)\vee \mathsf{Prepared}_T(CID)=1\big).
\end{aligned}
\]
When confidential execution is required, a deployment may further require an established trusted state:
\[
\mathsf{RequireConf}(CID)=true\Rightarrow trust\_flag=\textsf{trusted}.
\]

\subsection{Detailed OCI Entry-Point Rules}
\label{app:oci-rules}

All OCI entrypoints are evaluated against the persistent CID-level state record \(SR(CID)\). Updates must be linearizable and version-ordered. Repeated invocations must be idempotent and must not duplicate side effects.

\begin{itemize}
\item \texttt{create}(\(CID\)).
If \(SR(CID)\) does not exist, the runtime materializes the instance into its minimal persistent form, creates \(SR(CID)\) with \(S:=\textsf{Prepared}\), and initializes the host-side artifacts needed for subsequent multi-call behavior. If \(SR(CID)\) already exists, \texttt{create} is a no-op that returns success and leaves \(S\) unchanged. If materialization fails, \(SR(CID)\) remains absent and the instance stays in \textsf{Init}.

\item \texttt{start}(\(CID\)).
\texttt{start} is defined when \(SR(CID)\) exists and \(S\in\{\textsf{Prepared},\textsf{Running}\}\). If \(S=\textsf{Prepared}\), the runtime launches or reattaches to \(r_{\textsf{anchor}}\) and moves the state to \textsf{Running} after observing that the anchor is alive. If \(S=\textsf{Running}\), \texttt{start} is idempotent and returns success without changing \(S\). TEE-backed stages may remain inactive until requested by the anchor.

\item \texttt{state}(\(CID\)).
\texttt{state} returns the OCI projection \(\pi_{\textsf{OCI}}(S)\), together with auxiliary fields such as \(trust\_flag\), \(health\_flag\), \(tee\_phase\), and \(\mathsf{Ready}(CID)\). These fields do not affect the primary lifecycle state.

\item \texttt{wait}(\(CID\)).
\texttt{wait} blocks or polls until \(S\in\{\textsf{Stopped},\textsf{Failed}\}\), then returns the recorded composite exit information. Repeated \texttt{wait} calls observe the same terminal state and exit result.

\item \texttt{kill}(\(CID\)).
If \(S=\textsf{Running}\), the runtime terminates the REE anchor and cancels any ongoing protected stages. It then records \textsf{Stopped} for graceful termination or \textsf{Failed} when a failure condition is observed. If \(S\in\{\textsf{Stopped},\textsf{Failed}\}\), \texttt{kill} is a no-op.

\item \texttt{delete}(\(CID\)).
\texttt{delete} removes host-side artifacts and deletes \(SR(CID)\), returning the instance to \textsf{Init}. It is permitted only after terminal states under the adopted OCI contract. Repeated \texttt{delete} calls are idempotent.
\end{itemize}

At any time, if a failure of the REE anchor or any protected stage is observed,
\[
\mathsf{REEFail}(CID)\vee\mathsf{TEEFail}(CID),
\]
the runtime transitions the CID-level state to \textsf{Failed} and ensures that later \texttt{state}/\texttt{wait} calls reflect this terminal outcome.

\subsection{Composite Termination Semantics}
\label{app:termination}

OCI requires a single externally visible termination outcome, while a C4 instance may produce termination signals from the REE anchor, a TEE-backed stage, or a runtime policy controller. We model a termination-related event as
\[
E:=\langle src,\ code,\ reason\rangle,
\]
where \(src\in\{\textsf{R},\textsf{T},\textsf{P}\}\) denotes the REE anchor, TEE-backed stage, or policy/runtime controller, \(code\in\mathbb{Z}_{\ge 0}\) is an internal exit code, and
\[
reason\in\{\textsf{normal},\textsf{error},\textsf{untrusted},\textsf{killed},\textsf{policy}\}.
\]

When multiple events are observed, EBCC selects a dominant event according to:
\[
\textsf{untrusted}\succ\textsf{TEE-error}\succ\textsf{REE-error}\succ\textsf{killed}\succ\textsf{normal}.
\]
Let
\[
E^{\star}:=\max_{\succ}\mathcal{E}_{term}(CID)
\]
denote the selected event, with deterministic tie breaking. The externally visible exit code is then:
\[
exitCode:=\Phi(E^{\star}),
\]
where
\[
\Phi(E)=
\begin{cases}
c_{\textsf{untrusted}}, &
reason=\textsf{untrusted}\ \text{or}\ (src=\textsf{P}\wedge reason=\textsf{policy}),\\
code, & reason=\textsf{error},\\
0, & reason\in\{\textsf{normal},\textsf{killed}\}.
\end{cases}
\]
Thus security or policy violations are distinguishable from ordinary crashes, execution errors propagate their internal error code, and normal or user-requested termination maps to a successful stop.

For finite jobs, normal completion is anchored at the REE lifecycle anchor:
\[
\mathsf{Done}(CID)\Longleftrightarrow
(E^{\star}.src=\textsf{R})\wedge(E^{\star}.reason=\textsf{normal}),
\]
and all other terminal cases are failures:
\[
\mathsf{Fail}(CID)\Longleftrightarrow \neg\mathsf{Done}(CID).
\]
For service-mode workloads, termination may be triggered by user stop requests, policy revocation, failed trust checks, or failed health checks. These triggers are modeled as policy/runtime events with \(src=\textsf{P}\), allowing the same reduction function to produce a stable OCI-facing outcome.

\clearpage
\begin{landscape}

\section{Additional Model-to-TEE Mapping Details}
\label{app:mapping}

\begin{table}[H]
\centering
\caption{Full model-grounded mapping from EBCC transitions to backend-specific primitives (SGX/TDX/OP-TEE).}
\label{tab:full-mapping}
\vspace{2pt}

\footnotesize
\setlength{\tabcolsep}{6pt}
\renewcommand{\arraystretch}{1.15}

\resizebox{0.98\linewidth}{!}{%
\begin{tabular}{p{4.2cm} p{7.2cm} p{7.2cm} p{7.2cm}}
\toprule
\textbf{EBCC transition (model)} & \textbf{SGX (enclave)} & \textbf{TDX (TD/VM)} & \textbf{OP-TEE (TA)} \\
\midrule
\multicolumn{4}{l}{\emph{Instance-level (CID) lifecycle: TEE-agnostic OCI-managed anchor + persistent StateDir}}\\
\midrule
\texttt{create(CID)} &
\multicolumn{3}{p{21.6cm}}{Allocate \texttt{StateDir(CID)}; initialize \texttt{state.json}, request/response directories, EID sequence state, and artifact directories; perform idempotent update.}\\

\texttt{start(CID)} &
\multicolumn{3}{p{21.6cm}}{Launch the OCI-managed anchor; make the CID-visible runtime state enter the running phase. Backend-specific protected execution is not performed at this step.}\\

\texttt{wait(CID)} &
\multicolumn{3}{p{21.6cm}}{Observe anchor/container termination; persist exit status and final lifecycle state into \texttt{state.json}.}\\

\texttt{kill(CID)} &
Stop anchor; release SGX-side sessions or cached enclave handles if any &
Stop anchor; release TD-side RPC/vsock channel or runner session if any &
Stop anchor; close OP-TEE client sessions or shared-memory bindings if any \\

\texttt{delete(CID)} &
\multicolumn{3}{p{21.6cm}}{Remove \texttt{StateDir(CID)} and associated request, response, metadata, log, and evidence artifacts; perform idempotent cleanup.}\\

\midrule
\multicolumn{4}{l}{\emph{Stage-level (EID) pipeline: ReqPending $\rightarrow$ Claimed $\rightarrow$ Prepared $\rightarrow$ Executing $\rightarrow$ Completed/Failed}}\\
\midrule

\texttt{ReqPending} &
\multicolumn{3}{p{21.6cm}}{Anchor emits a TEE-agnostic stage request, e.g., \texttt{stage\_req\_*}, including the stage name, request identifier, nonce, response path, and authentication fields.}\\

\texttt{Claimed} &
\multicolumn{3}{p{21.6cm}}{Runtime atomically claims the request; checks freshness and request metadata; allocates a fresh EID; creates \texttt{enclaves/<EID>/}; initializes per-stage metadata.}\\

\texttt{Prepared} &
Create or reuse an enclave context; bind enclave measurement/session information to the EID &
Ensure the TD-side runner/channel is ready; bind the RPC/vsock session, channel identifier, and measurement/report context to the EID &
Open or reuse a TA session via \texttt{TEEC\_OpenSession}; bind TA UUID, command identifier, and shared-memory/session state to the EID \\

\texttt{Executing} &
Enter the enclave through \texttt{ECALL}; execute the selected stage; return rc/output and optional quote or measurement evidence &
Send an RPC/vsock request into the TD; execute the selected stage inside the confidential VM; return rc/output and optional TD report evidence &
Invoke the TA command through \texttt{TEEC\_InvokeCommand}; execute the selected TA operation; return rc/output and TA-side result metadata \\

\texttt{Completed/Failed} &
Persist \texttt{run.log}; update \texttt{meta.json} with \texttt{tee=sgx}, EID, stage, rc, timing, measurement/quote fields if available; write \texttt{.resp} &
Persist \texttt{run.log}; update \texttt{meta.json} with \texttt{tee=tdx}, EID, stage, rc, timing, channel/session, measurement/report fields if available; write \texttt{.resp} &
Persist \texttt{run.log}; update \texttt{meta.json} with \texttt{tee=optee}, EID, stage, rc, timing, TA UUID/session/cmd fields; write \texttt{.resp} \\

\bottomrule
\end{tabular}%
}
\end{table}
\end{landscape}

\end{document}